\documentclass[preprint,pteplogo]{ptephy_v2}%%%%%% to generate preprint number with ptep logo
 
\preprintnumber{XXXX-XXXX} %%% %%% Insert preprint number here
\usepackage{hyperref}
\usepackage{amssymb}
\usepackage{amsthm}
\usepackage{bm}
\usepackage{amsmath}
\usepackage{tabularx}
\usepackage{lineno}
\usepackage{boxedminipage}
\usepackage{color}
\usepackage{url}
\usepackage{hyperref}
\usepackage{ulem}
\usepackage{multirow}
\usepackage{arydshln}
\usepackage{subfigure}
\usepackage{diagbox}
\usepackage{lscape}	% required for `\landscape' (yatex added)

\begin{document}

%%%\begin{frontmatter} %% comout 2026/4/15
%\title{Upgrade of Honda atmospheric neutrino flux calculation with implementing recent hadron interaction measurement}
\title{Low-energy atmospheric neutrino flux calculation with accelerator-data-driven tuning}

\author[1]{Kazufumi Sato}
\author[2,3]{Hiroaki Menjo}
\author[1,2,3]{Yoshitaka Itow}
\author[1]{Morihiro Honda}
\affil[1]{Institute for Cosmic Ray Research, the University of Tokyo, Kashiwa, Chiba, Japan \email{kazufumi@km.icrr.u-tokyo.ac.jp}}
\affil[2]{Institute for Space-Earth Environmental Research, Nagoya University, Nagoya, Aichi, Japan}
\affil[3]{Kobayashi-Maskawa Institute for the Origin of Particles and the Universe, Nagoya University, Nagoya, Aichi, Japan}

%% comout 2026/4/15
% \author[ICRR]{Kazufumi Sato}
% \author[ISEE,KMI]{Hiroaki Menjo}
% \author[ICRR,ISEE,KMI]{Yoshitaka Itow}
% \author[ICRR]{Morihiro Honda}

% \cortext[cor1]{{\it E-mail address:}kazufumi@km.icrr.u-tokyo.ac.jp }

% \address[ICRR]{Institute for Cosmic Ray Research, the University of Tokyo, \\
% Kashiwa, Chiba, Japan}

% \address[ISEE]{Institute for Space-Earth Environmental Research, Nagoya University, \\ 
% Nagoya, Aichi, Japan}

% \address[KMI]{Kobayashi-Maskawa Institute for the Origin of Particles and the Universe, Nagoya University, \\
% Nagoya, Aichi, Japan}

\begin{abstract}

 We have incorporated a hadron interaction tuning based on accelerator data into our atmospheric neutrino flux calculation, which has been used to analyze atmospheric neutrino oscillations at Super-Kamiokande. This new approach enables a more direct evaluation of the flux uncertainty than a conventional tuning using atmospheric muons. 
 The neutrino flux calculated with this new tuning is 5\%--10\% smaller but still consistent with our previously published prediction within its uncertainty. The flavor ratio $(\nu_{\mu}+\bar{\nu}_{\mu})/(\nu_e+\bar{\nu}_e)$ and  $\bar{\nu}/\nu$ ratios were consistent with the previous prediction. Based on the measurement errors of the accelerator data, we evaluated the flux uncertainty associated with the new tuning to be 7\%--9\% in the $E_{\nu} <$ 1 GeV region, which was difficult to assess with the conventional tuning. The flux uncertainty in the $1<E_{\nu}<10$ GeV region was evaluated to be 5\%--7\%,  which is an improvement over the conventional tuning. 

\end{abstract}

\subjectindex{cosmic ray, atmospheric neutrino, simulation}

\maketitle

% \begin{keyword}
% cosmic ray, atmospheric neutrino, simulation
% \end{keyword}

%% comout 2026/4/15
%\end{frontmatter}

%\linenumbers

\section{Introduction}
Atmospheric neutrinos are produced through the decays of particles in air showers, which are cascades of hadronic and electromagnetic interactions caused by the collision of high-energy cosmic rays with Earth's atmosphere. The atmospheric neutrinos span wide ranges of energy ($O$(10 MeV)--$O$(PeV)) and flight length ($O(10)$--$O(10^4)$ km) and are powerful tools for studying various physics topics, including neutrino oscillations.

Accurate prediction of neutrino flux is essential for experimental research on atmospheric neutrinos. Our calculation using three-dimensional (3D) Monte Carlo simulations (MC) of air showers~\cite{HKKM2011,HKKM2015}  provides flux prediction in the energy region of 0.1--$10^4$ GeV, and it has been used in several neutrino experiments, including Super-Kamiokande~\cite{SK}.

The main uncertainty in the neutrino flux prediction results from hadron interactions in the air shower. Previously, we tuned the hadronic interaction model in the MC based on atmospheric muon flux observations~\cite{BESS}. This ``$\mu$-tuning''~\cite{MuTune} suppresses the flux uncertainty down to $\sim$7\% in the $1 < E_{\nu} < 10$ GeV region~\cite{HKKM2006}. However, relatively large uncertainties exist at $E_{\nu}<1$ GeV and $E_{\nu}> 10$ GeV. The former is because low-energy muons mostly decay before reaching the detector at ground level. The latter is because kaon decays contribute to neutrino production in this energy region while muons are mostly produced from pion decays. 

Reducing the uncertainty of low-energy atmospheric neutrino flux is important for various physics topics. Neutrinos around 0.2--1 GeV can be a main background source of diffuse supernova neutrino background (DSNB) searches~\cite{DSNB}, and $O$(0.1 GeV) neutrinos produce an ultimate background of direct dark matter searches known as "neutrino fog"~\cite{NeutrinoFog}. A nonzero CP phase in the lepton sector is predicted to enhance/decrease the neutrino flux around 0.2--1 GeV through atmospheric neutrino oscillation~\cite{HKDR}. Oscillation at 2--10 GeV shows enhancement of $\nu_{\mu}\to\nu_{e}$ or $\bar{\nu}_{\mu}\to\bar{\nu}_{e}$, depending on whether the mass hierarchy is normal or inverted~\cite{HKDR}. Thus, a new tuning method, different from the conventional $\mu$ tuning, is desirable to reduce the uncertainty in the low-energy region, from $O(0.1)$ GeV to 10 GeV. %Thus, a new tuning method different from the conventional $\mu$ tuning is desirable to reduce the uncertainty in low energy region from $O(0.1$ GeV) to 10 GeV.

In this study, we tuned the hadron interaction model in the MC based on hadron production data measured in accelerator experiments. This new method is expected to improve the uncertainty in low-energy regions because the accelerator data cover large phase space relevant to low-energy neutrino production, as discussed later. This approach is also beneficial for oscillation analyses combined with long-baseline experiments, where beam production simulations are tuned using accelerator data. To isolate the impact of the new tuning method, we used the same models for all physics inputs other than hadron production as in the previous calculation~\cite{HKKM2015}, including electromagnetic interactions,  the Earth's magnetic field, air density, and primary cosmic ray flux. %To observe the impact of the tuning method, this study used the same models for physics other than hadron production as in previous calculations~\cite{HKKM2015}, such as electromagnetic interactions,

We briefly review our flux simulation in Section~\ref{sec:hadr-inter-honda}, particularly the treatment of the hadron interaction in the simulation code. In Section~\ref{sec:overv-accel-data}, our accelerator-data-driven tuning method is overviewed. The accelerator data used in our study are listed in Section~\ref{sec:accelerator-data}. Their coverage of phase space relevant to neutrino production is evaluated in the same section. We develop the parameterization to describe the accelerator data in Section~\ref{sec:parameterization}. The tuning result is shown in Section~\ref{sec:accel-data-driv-2}, and systematic uncertainties related to the tuning are evaluated in Section~\ref{sec:syst-uncert-1}. Section~\ref{sec:summary} concludes our paper. 

\section{Simulation configuration}
\label{sec:hadr-inter-honda}
\begin{figure}[!th]
\centering
\includegraphics[width=0.98\textwidth]{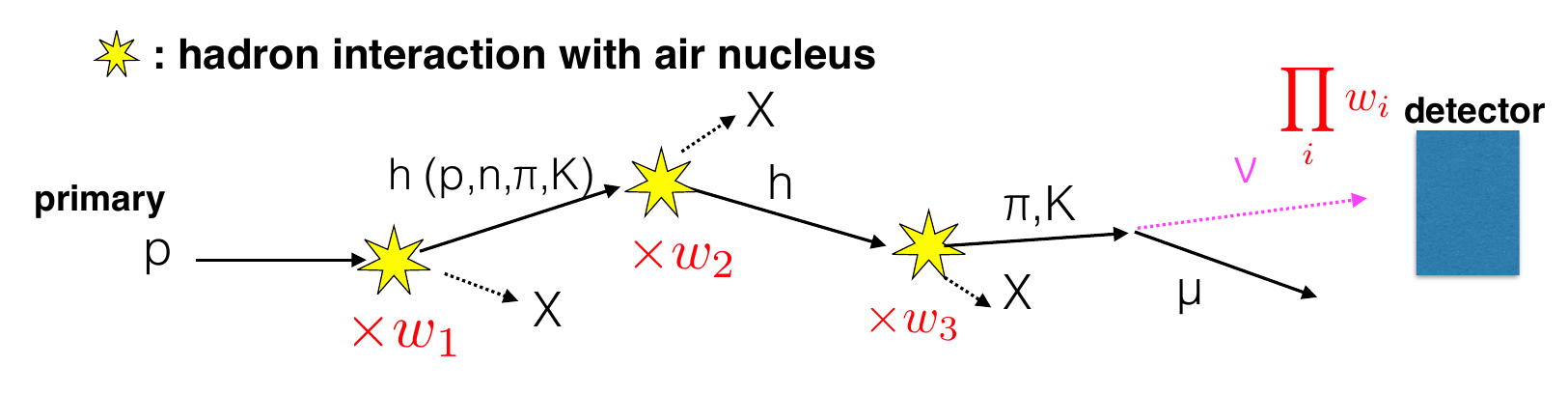}
\caption{Schematic view of chain interactions associated with neutrino production. The $w_i$ represents a weight for our tuning (Eq.~\ref{eq:weight}) at the $i$-th vertex. }
 \label{fig:interactionChain}
\end{figure}

Our simulation \cite{HKKM2015} is a full 3D MC that tracks all particles produced in the air shower. 
In the simulation, primary cosmic rays (protons and nuclei) are injected from 100 km altitude into the Earth's atmosphere with the flux model~\cite{HKKM2004} based on measurements by satellite and balloon experiments~\cite{AMS,BESS}. The particles are transported in the air with a geomagnetic field based on the IGRF model~\cite{IGRF} and then interact with a nucleus in the air in accordance with atmospheric density modeled by NRMSISE-00~\cite{NRMSISE}. All secondary particles are also tracked in the same manner. As illustrated in Fig.~\ref{fig:interactionChain}, neutrinos are eventually produced through particle decays at the end of a chain of hadronic interactions. Among these neutrinos, the energy and arrival direction of those hitting a virtual detector placed at an arbitrary location are recorded.

Particle production from hadronic interactions is simulated based on the event generator JAM~\cite{JAM} for incident kinetic energies below 31 GeV/c and on DPMJET-III~\cite{DPM} for higher energies.
Instead of running the full event generator, we use an inclusive approach based on pre-tabulated multiplicity and momentum distributions, which simplifies and speeds up the calculation. The multiplicity of each particle species is given by a lookup table: 
\begin{equation}
    f_{mul}(E_{kin},x_{in},x_{out}),
\end{equation}
where $E_{kin},x_{in}$, and $x_{out}$ are input parameters of the table, which represent the kinetic energy of an incident particle, a type of incident particle, and a type of generated particle, respectively. The following light and long-lived hadrons are considered as $x_{in}$ and $x_{out}$: $x_{in(out)} = p,n,\bar{p},\bar{n},\pi^\pm,K^\pm,K^0$. %As for $x_{out}$,  electromagnetic cascade particles ($e^\pm,\gamma$,  $\pi^0$) are considered in addition. 
Other hadrons are considered to decay immediately and only their decay products appear in the simulation. When the incident particle is He, it is decomposed into two protons and two neutrons, and then the $f_{mul}$ for nucleons is applied. Thus, from an interaction caused by $x_{in}$ with $E_{kin}$ energy, we immediately calculate how many $\pi^{+}$s, $\pi^{-}$s, $K^{+}$s, ... are generated. If $f_{mul}$ is not an integer, it is stochastically rounded (e.g., if $f_{mul}$ = 3.2, the value 3 (4) is selected with a probability of 0.8 (0.2).)  % If $f_{mul}$ is not an integer, an integer before or after $f_{mul}$ is randomly selected based on the digits after the decimal point of $f_{mul}$.
In the code, the calculation of the momentum and direction, $p_{out}$ and $\theta$, of each generated particle is also sped up by random sampling from the pre-defined $p$-$\theta$ distribution, 
\begin{equation}
f_{p\textrm{-}\theta}\left(E_{kin},x_{in},x_{out}; p_{out},\theta\right).
\end{equation}
Similar to $f_{mul}$, the $f_{p\textrm{-}\theta}$ also takes $E_{kin},x_{in}$, and $x_{out}$ as input parameters. The parameterization of $f_{p\textrm{-}\theta}$ is written in \cite{HKKM2011}. Although the inclusive code cannot simulate the correlation between secondary particles, it correctly reproduces the kinematics relation between the incident particle and each secondary particle. This is sufficient for our purpose, as long as correlations between two neutrinos produced in the same air shower are not required. %It is sufficient for our purpose unless we simulate a correlation of two neutrinos. 

$f_{mul}$ and $f_{p\textrm{-}\theta}$ were constructed by simulating collisions of a hadron with an air nucleus using the original JAM and DPMJET-III models. Collisions of a hadron with air-nuclei were simulated repeatedly, for each type of incident hadron  $x_{in}$ and for various injection energies ($E_{kin} = 10^{-1},10^{-0.9},...,10^{1.5}$ GeV with JAM and $E_{kin} = 10^{1.5},10^{2},10^{2.5},...,10^{6}$ with DPMJET-III). The average number of the generated particles was used as $f_{mul}$, and the accumulated and normalized $p$-$\theta$ distribution was used to reconstruct $f_{p-\theta}$. %The results of N and O were averaged based on the air composition.
%To construct $f_{mul}$ and $f_{p\textrm{-}\theta}$,  collisions of a hadron with an air nucleus were simulated repeatedly based on the original JAM and DPMJET-III models, for each kind of incident hadron $x_{in}$ and for each injection energy  ($E_{kin} = 10^{-1},10^{-0.9},...,10^{1.5}$ GeV with JAM and $E_{kin} = 10^{1.5},10^{2},10^{2.5},...,10^{6}$ with DPMJET-III).   $p_{out}$ and $\theta$ of each secondary particle generated from the collision were recorded. The average number of the generated particles is used as $f_{mul}$, and the accumulated and normalized $p-\theta$ distribution is used to reconstruct $f_{p-\theta}$. For the case that the incident particle is He, it is decomposed into 2 protons and 2 neutrons, and then the $f_{mul}$ and $f_{p\textrm{-}\theta}$ for nucleon are applied. 

For the conventional $\mu$-tuning, the $f_{p\mathrm{-}\theta}$ distribution was modified~\cite{HKKM2006,HKKM2011} to reproduce atmospheric muon observations~\cite{BESS}. The accelerator-data-driven tuning introduced in this paper modifies the product $f_{mul} \times f_{p-\theta}$, as described in the next section.

\section{Overview of accelerator-data-driven tuning}
\label{sec:overv-accel-data}

Several accelerator experiments~\cite{HARP,HARP_LA,HARP_proton,E910,NA61,NA49,BMPT} present the measurements of inclusive differential cross-sections of hadron interaction. Our aim in this paper is to tune the simulation based on these measurements. 

Before describing the tuning in detail, we first unify the notation. To represent the differential cross-section, some of the experiments use $\frac{d^2\sigma}{dpd\theta}$, whereas others use $\frac{d^2\sigma}{dpd\Omega}$, $\frac{d^2\sigma}{dx_{F}dp_{T}}$, etc.  Additionally, the notation of the momentum of the secondary particle differs depending on the experiments; some use $p_{out}$ and $\theta$, whereas others use transverse momentum $p_T$ and Feynman-X $x_{F}$, which is defined as 
\begin{equation}
 x_F \equiv \frac{p_{L}}{p_\textrm{max}},
\end{equation}
where $p_{L}$ is the longitudinal momentum of a secondary particle in the CM frame of a nucleon-nucleon collision, and $p_\textrm{max}$ is the theoretical maximum momentum. Hereafter, we use $x_{F}$ and $p_T$ to represent the kinematics of the secondary particle, and we use an invariant form of the differential cross-section:
\begin{equation}
 E\frac{d^3\sigma}{dp^3}(p_{in},x_{in},A,x_{out},x_{F},p_T),
\end{equation}
where $p_{in}$ is the incident particle momentum (= beam momentum), and $A$ is the atomic mass number of the target atom. The differential cross-sections reported in the form of $\frac{d^2\sigma}{dpd\theta}$, $\frac{d^2\sigma}{dpd\Omega}$, $\frac{d^2\sigma}{dx_{F}dp_{T}}$, ..., are appropriately transformed into $E\frac{d^3\sigma}{dp^3}$.

For the comparison of the measured $E\frac{d^3\sigma}{dp^3}$ with the MC, we introduce the differential form of multiplicity $E\frac{d^3n}{dp^3}$, which is defined as %the number density defined as  
\begin{equation}
    E\frac{d^3n}{dp^3}(p_{in},x_{in},A,x_{out},x_{F},p_T) \equiv \frac{1}{\sigma_{prod}\left(p_{in},x_{in}\right)} \times E\frac{d^3\sigma}{dp^3}(p_{in},x_{in},A,x_{out},x_{F},p_T)
\end{equation}
where $\sigma_{prod}$ is a production cross-section, that is, a cross-section to produce at least one hadron. $E\frac{d^3n}{dp^3}$ represents the number density of $x_{out}$ particles generated into the infinitesimal phase space $d^3p/E$ at ($x_F,p_T$), from the collision of $x_{in} + A$ with the incident momentum $p_{in}$. As introduced in the previous section, we use $f_{mul}$ and $f_{p-\theta}$ in the simulation. From the definition, the product of these quantities represents the number density in $p$-$\theta$ plane: 
\begin{equation}
f_{p\textrm{-}\theta} \times f_{mul} = \frac{1}{\sigma_{prod}}\frac{d^2 \sigma}{dpd\theta}\left(p_{in},x_{in},A_{air},x_{out},p_{out},\theta \right),
\end{equation}
where $A_{air} = 14.5$ is the average mass number of nuclei in the atmosphere.  By changing to the invariant form and using $x_F$ and $p_T$ instead of $p_{out}$ and $\theta$, we obtain $E\frac{d^3n}{dp^3}$ from $f_{mul} \times f_{p\textrm{-}\theta}$. Thus, we can directly compare $E\frac{d^3\sigma}{dp^3}$ measured in the experiments with $E\frac{d^3 n}{dp^3}$ implemented in the simulation. We define a weight $w$ to describe the difference between the data and simulation as
\begin{equation}
\label{eq:weight}
 w(p_{in},x_{in},x_{out},x_F,p_T) \equiv \frac{ [ E\frac{d^3\sigma}{dp^3}(p_{in},x_{in},A_{air},x_{out},x_{F},p_T)]_{data} } { [\sigma_{prod} E\frac{d^3n}{dp^3}(p_{in},x_{in},A_{air},x_{out},x_{F},p_T)]_{MC}},
\end{equation}
where the subscripts “Data” and “MC” indicate experimentally measured and simulated values, respectively. $w$ was prepared for each $x_{in}$ and $x_{out}$ value and is a continuous function of $p_{in},x_{F}$, and $p_T$. In the following sections, details of the derivations of the $[ E\frac{d^3\sigma}{dp^3}]_{data}$ and $w$ are described. 

The accelerator-data-driven tuning was performed by applying the weights to hadron interactions in the simulation. First, we ran our simulation with conventional hadron production model. In the simulation, atmospheric neutrinos are produced through a chain of hadronic interactions in the air shower cascade, as illustrated in Fig.~\ref{fig:interactionChain}. We recorded $p_{in},x_{in},x_{out},x_{F},$ and $p_{T}$ at every hadron vertex on the chain and applied the weight $w(p_{in},x_{in},x_{out},x_F,p_T)$ to each vertex. The product $W_{event} \equiv \prod_i w_i$ was used as an event weight when counting the number of neutrinos recorded in a detector, where $w_i$ represents the weight applied to the $i$-th vertex on the chain. Thus, the flux is tuned to reflect the difference between the data and MC. %Subsequently, atmospheric neutrinos were produced in the air shower cascade as a result of a chain of interactions, as indicated in Fig.~\ref{fig:interactionChain}. 

\section{Accelerator data}
\label{sec:accelerator-data}
In this and the next two sections, we derive the weighting function defined in Eq.~(\ref{eq:weight}). %we develop the weighting function shown in Eq.~\ref{eq:weight}.

First, in this section, we select the accelerator data to derive  $[E\frac{d^3\sigma}{dp^3}]_{data}$ in Eq.~\ref{eq:weight} and evaluate whether these data cover sufficient phase space for our study. We referred to several fixed-target accelerator data: HARP~\cite{HARP,HARP_LA,HARP_proton}, BNL E910~\cite{E910}, NA61~\cite{NA61}, NA49~\cite{NA49}, NA56/SPY, and NA20~\cite{BMPT}. These experiments use a proton beam whose momentum ranges from 3 to 450 GeV/$c$, and they provided inclusive differential cross-sections of $\pi^{\pm}, K^{\pm},$ and/or proton productions, as summarized in Table~\ref{tab:usedData}.

\begin{table*}
  \caption{List of accelerator data used for the tuning. These data provide the inclusive differential cross-section for the interaction $p + A \to x_{out} + X$. The types of target atoms and the reference number are shown in each cell.}
 \label{tab:usedData}
   
   \begin{tabular}{c | c | c | c | c | c | c |}
    &  \multicolumn{6}{c|}{Beam momentum [GeV/c]} \\
    \cline{2-7}
    $x_{out}$  & 3 & 5 & 6.4 & 8 & 12 & 12.3 \\
    \hline 
    $\pi^{\pm}$  & Be, C, Al  &  Be, C, Al  & Be & Be, C, Al  & Be, C, Al  & Be \\
    & \cite{HARP,HARP_LA}  &  \cite{HARP,HARP_LA}  & \cite{E910} & \cite{HARP,HARP_LA}  & \cite{HARP,HARP_LA}  & \cite{E910} \\
    \hline 
    $K^{\pm}$  & --  & -- &--  &--  &--  &-- \\
    &  &    &  &   &   &  \\
    \hline 
    $p$  & Be, C, Al  &  Be, C, Al  & -- & Be, C, Al  & Be, C, Al  &-- \\
    & \cite{HARP_proton}  &  \cite{HARP_proton}  &  & \cite{HARP_proton}  & \cite{HARP_proton}  & \\
    \hline 
   \end{tabular}
\vskip.5\baselineskip
   \begin{tabular}{c | c | c | c | c | c | }
      &  \multicolumn{5}{c|}{Beam momentum [GeV/c]} \\
    \cline{2-6}
    $x_{out}$  & 17.5 & 31 & 158 & 400 & 450 \\ 
    \hline 
    $\pi^{\pm}$  & Be & C & C & Be & Be \\ 
    & \cite{E910} & \cite{NA61} & \cite{NA49} & \cite{BMPT} & \cite{BMPT} \\ 
    \hline 
    $K^{\pm}$  & -- & C & -- & Be & Be \\ 
    &   & \cite{NA61} &  & \cite{BMPT} & \cite{BMPT} \\ 
    \hline 
    $p$  & --  & C &C  & Be & Be \\ 
    &  & \cite{NA61} & \cite{NA49} & \cite{BMPT} & \cite{BMPT} \\ 
    \hline 
   \end{tabular}

%  \end{center}
 \end{table*}

\begin{figure}[!th]
\centering
\includegraphics[width=0.98\textwidth]{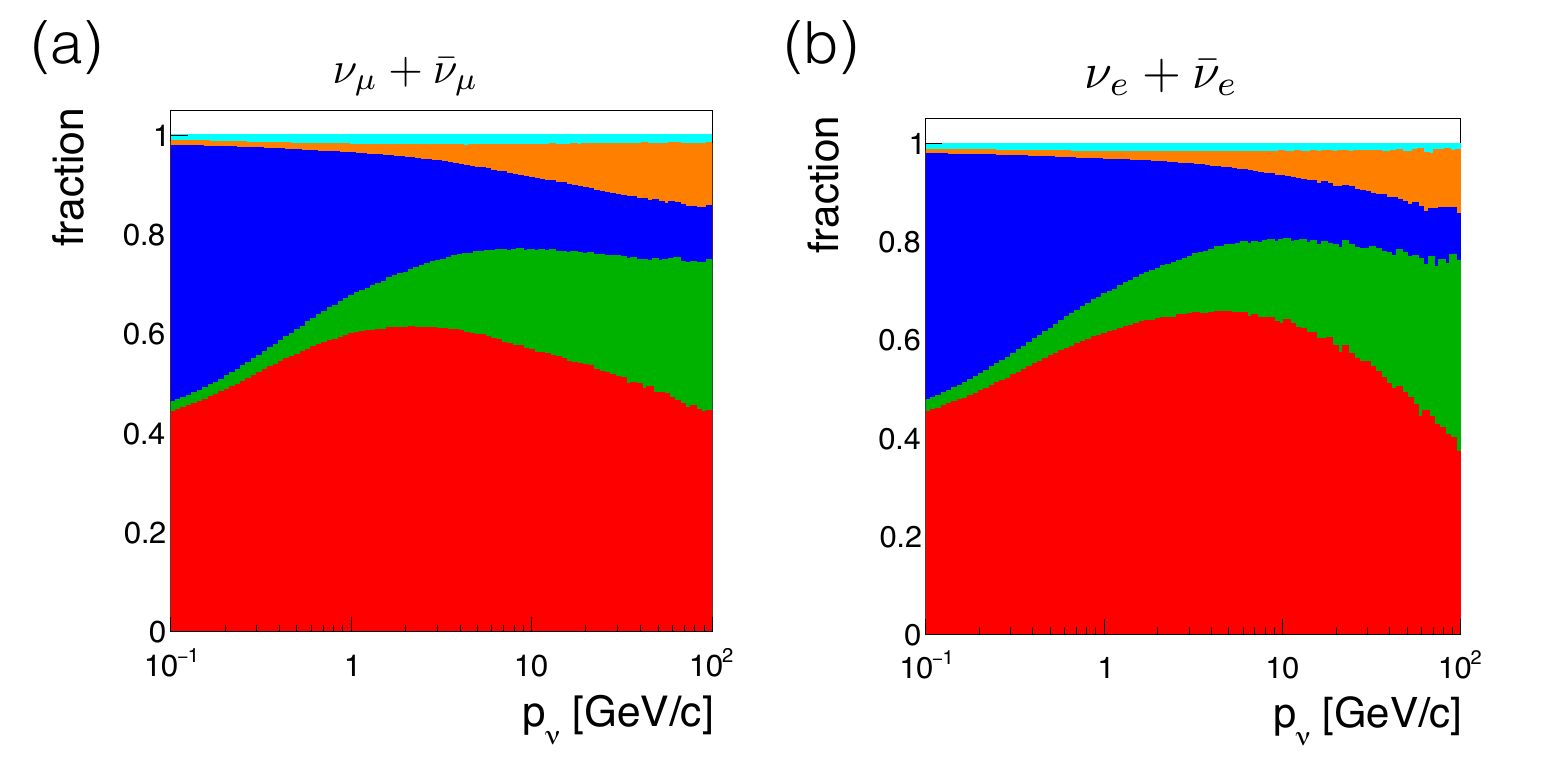}
\caption{Hadron interaction types associated with $\nu_{\mu}$ or $\bar{\nu}_{\mu}$ production (a) and  $\nu_{e}$ or $\bar{\nu}_{e}$ production (b). Histograms are normalized by total number of hadron interactions related to neutrino production.%, except the ones caused by alpha particles.
 Colors show combination of $x_{in}$ and $x_{out}$ of $x_{in} + \textrm{Air} \to x_{out} + X$ interaction. In the red, green, and blue histograms, $x_{in}$ is for the nucleon and $x_{out}$ is for $\pi^{\pm}, K$, and nucleon, respectively. The orange histogram shows collisions where $x_{in} = $ meson. The cyan shows interaction where $x_{in}$ and/or $x_{out}$ are anti-nucleons.}
\label{fig:typesOfInteraction}
\end{figure}

Our first concern was whether these measurements provide sufficient data to cover the phase space relevant to the neutrino productions.  We simulated the atmospheric neutrino production using our MC code to evaluate that. 
As illustrated in Fig.~\ref{fig:interactionChain}, neutrinos are produced after chains of interactions including hadron collision vertexes with air: $x_{in} + Air \to x_{out} + X$, where $x_{in}$ and $x_{out}$ represent incident and secondary hadrons, respectively. We classified the hadron vertexes into five categories according to the combination of $x_{in}$ and $x_{out}$; $\left(x_{in}\to x_{out}\right)$ = $\left(p,n\to \pi^{\pm}\right)$, $\left(p,n\to K^0,K^\pm\right)$, $\left(p,n\to p,n\right)$, $\left(\pi,K\to any\right)$, and others. Interactions caused by He were decomposed to nucleon-induced interactions. 
The breakdown of the interaction categories is shown in Fig.~\ref{fig:typesOfInteraction} as a function of the neutrino momentum $p_{\nu}$. %In $p_{\nu}<100$ GeV/$c$ 
We focus on neutrinos below 10 GeV, and in that energy region, hadron interactions are primarily caused by incident nucleons while the interactions caused by mesons are not dominant. Therefore, we decided to use only proton-beam data in this study. Pion production is the leading process in the $p_{\nu} > 1$ GeV/$c$ region, whereas nucleon production is dominant for lower $p_{\nu}$. The contribution of kaon production is small but not negligible in the $p_{\nu} <$ 10 GeV/$c$ region. %The accelerator experiments in Table~\ref{tab:usedData} provide $\pi,K,p$ productions from proton beam, which cover the main contributions of low-energy neutrino production. 

\begin{figure}[!th]
\centering
\includegraphics[width=0.98\textwidth]{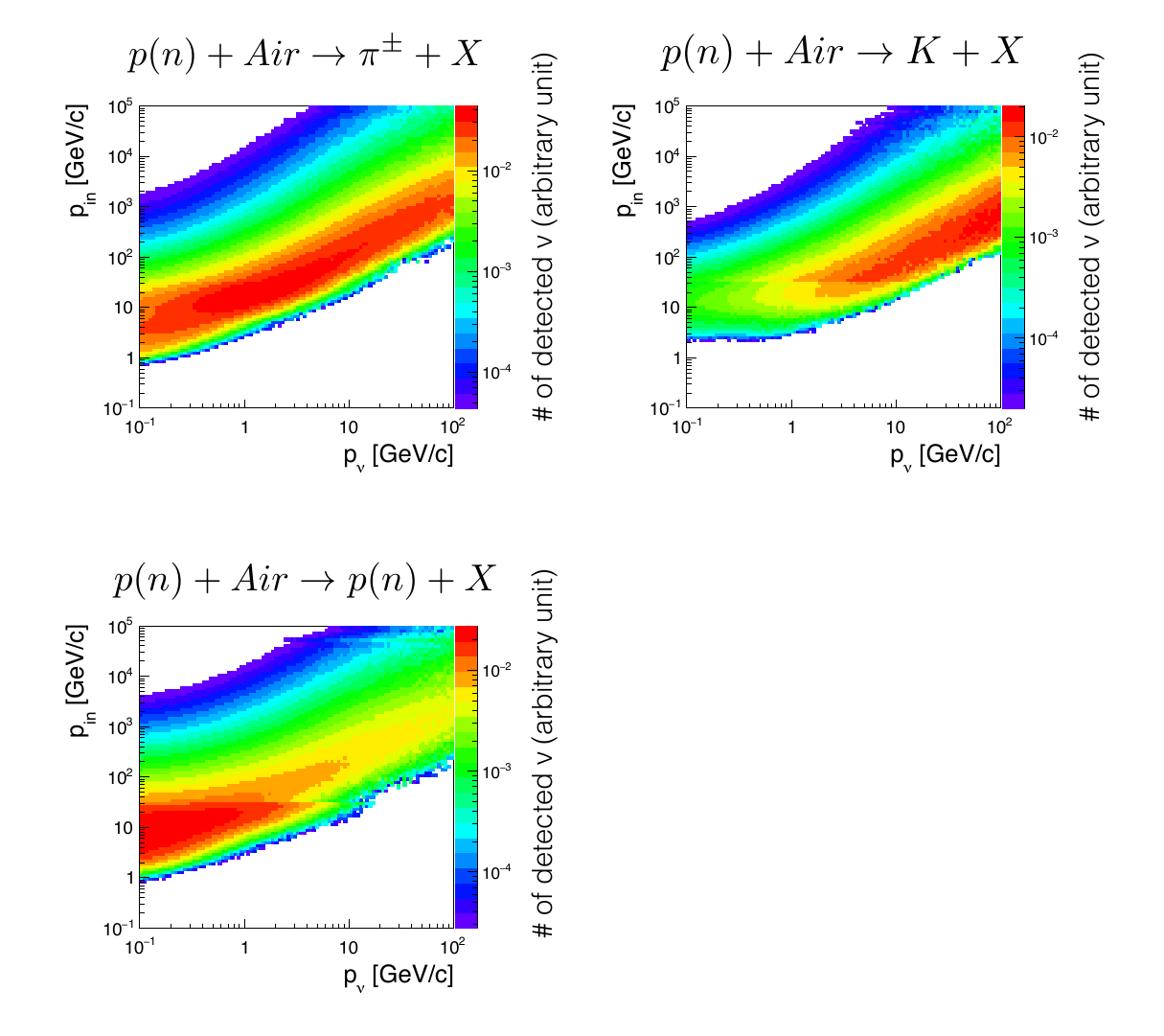}
\caption{Momentum distribution of incident particle in hadron interaction associated with $\nu_{\mu}$ or $\bar{\nu}_{\mu}$ production as a function of neutrino momentum. The left, center, and right plots show interactions where outgoing particles are $\pi^\pm, K$, and nucleon, respectively.}
\label{fig:pin_vs_pnu}
\end{figure}

We also checked the relation between the neutrino momentum $p_{\nu}$ and incident particle momentum $p_{in}$, as shown in Fig.~\ref{fig:pin_vs_pnu}.  If $p_{\nu} < 100$ GeV/$c$, the peak of $p_{in}$ falls within the range from $p_{in}$ = 3 to 450 GeV/$c$, which are the minimum and maximum beam momenta of the data listed in Table~\ref{tab:usedData}.

\begin{figure}[!th]
\centering
\includegraphics[width=0.5\textwidth]{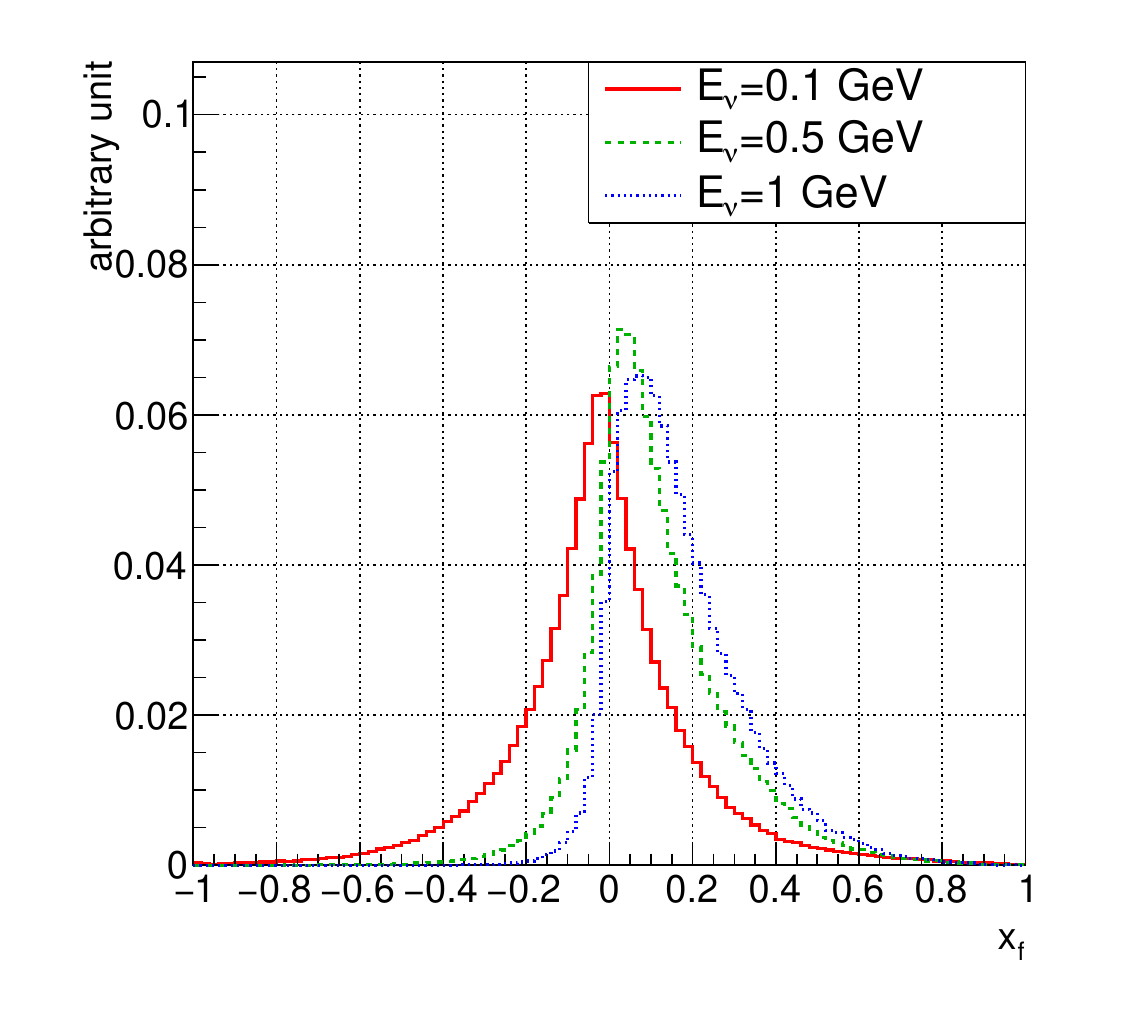}
\caption{Feynman-X $x_{f}$ distribution of the parent mesons of neutrinos.}
\label{fig:xf1d}
\end{figure}

\begin{figure}[!th]
 \centering
 \includegraphics[width=0.95\textwidth]{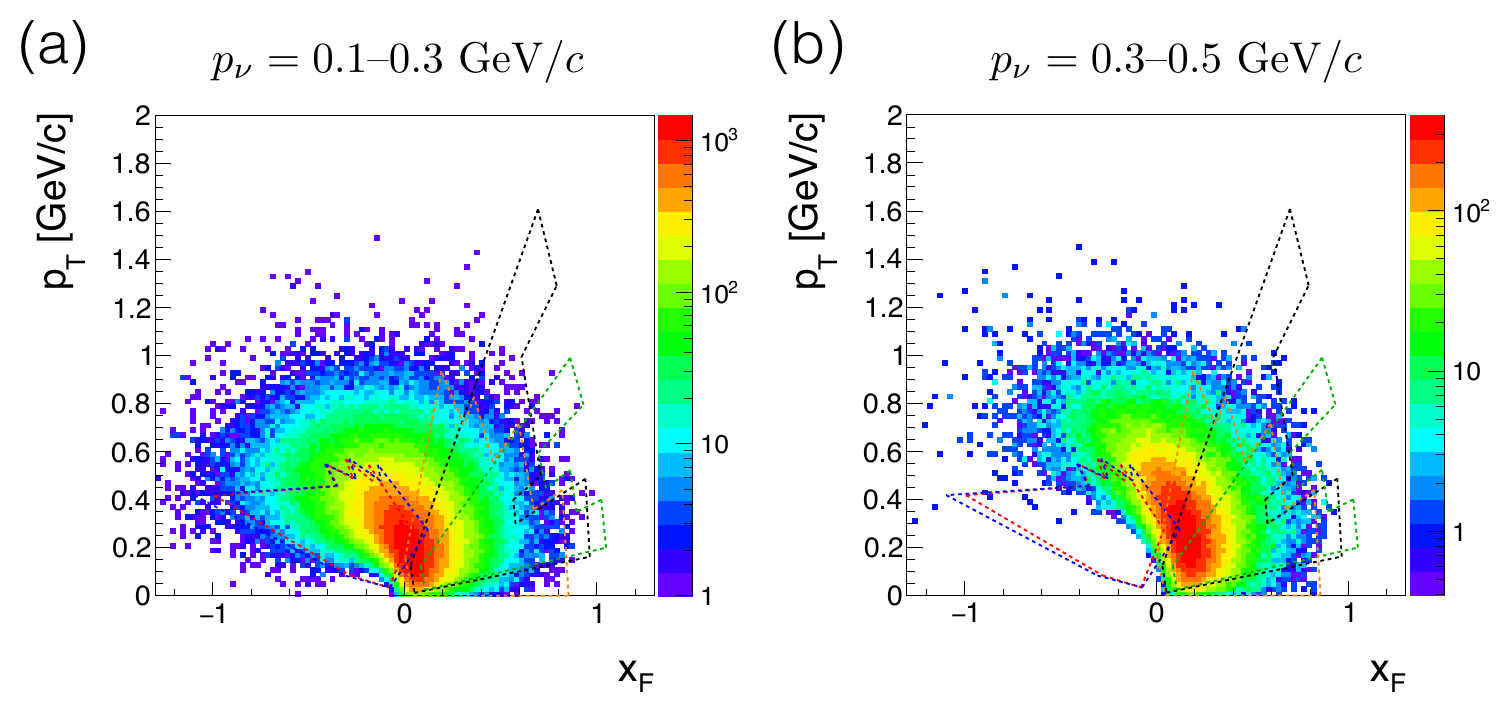}
 \caption{Example of phase-space coverage in $x_{F}$--$p_{T}$ plane. The 2-D histogram shows phase-space of $\pi^{\pm}$ production associated with 0.3--0.5 GeV/$c$ $\nu_{\mu}$, where the momentum of incident particle is selected to be $p_{in} = 3$--$5$ GeV/$c$. The regions surrounded by magenta and cyan lines are covered by HARP data with beam momentum = 3 or 5 GeV/$c$.}
 \label{fig:exampleOfCoverage}
\end{figure}

\begin{figure}[!th]
 \centering
 \includegraphics[width=0.95\textwidth]{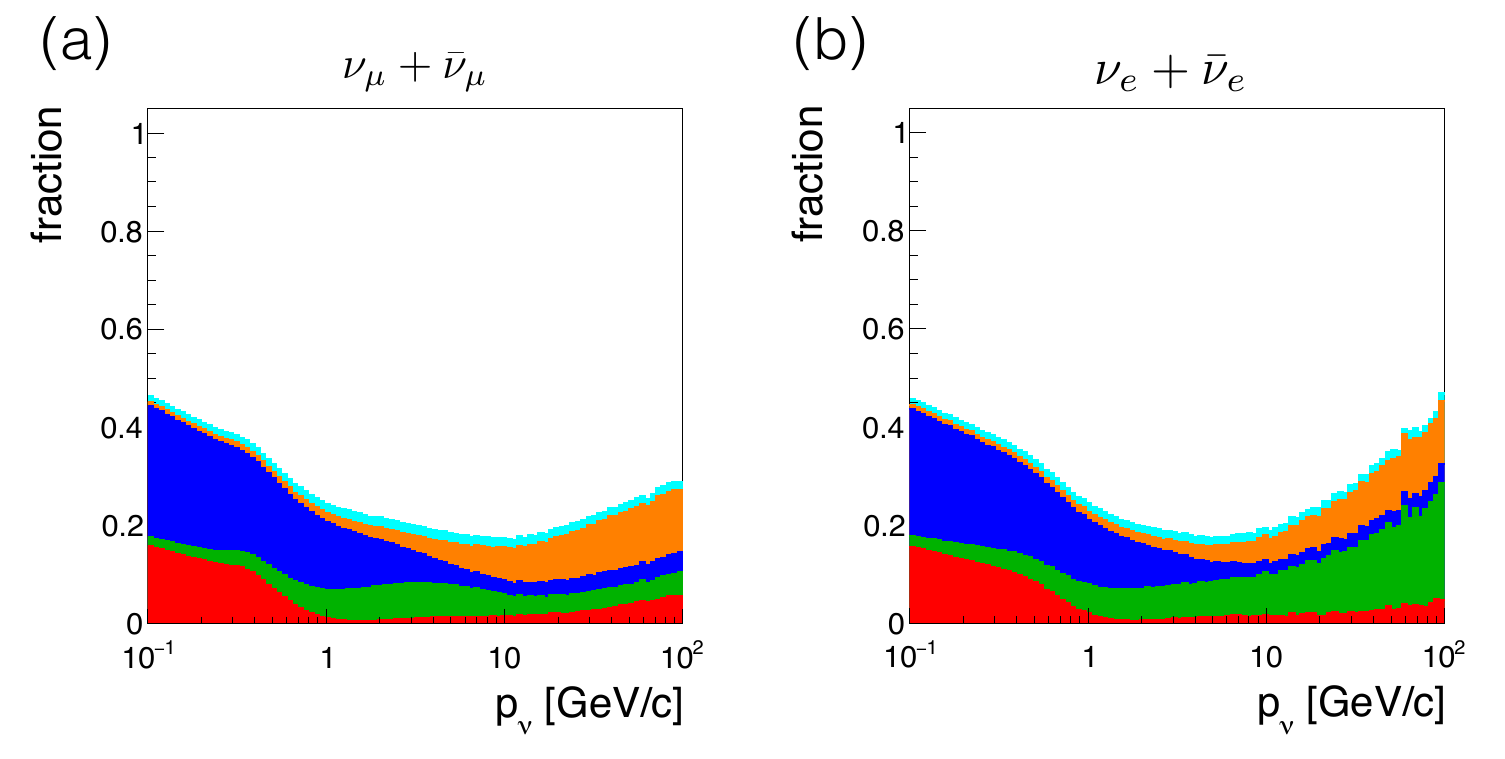}
 \caption{Fraction of phase-space not covered by any beam data,  related to $\nu_\mu$ and $\bar{\nu}_\mu$ production (a) and $\nu_e$ and $\bar{\nu}_e$ production (b). Histograms are normalized by total number of hadron interactions related to neutrino production.
The color meanings are the same as in Fig.~\ref{fig:typesOfInteraction}.}
 \label{fig:coverage}
\end{figure}

One of the features of the low-energy region is that secondary mesons with small $x_{F}$ are involved in neutrino production. Figure~\ref{fig:xf1d} shows the expected $x_{F}$ distribution of the parent mesons of neutrinos for several $p_{\nu}$. Only mesons with positive $x_{F}$, that is, mesons flying in the forward direction, contribute to the production of neutrinos with energies above 1 GeV. In contrast, in the energy region well below 1 GeV, significant neutrinos come from mesons with negative $x_F$, that is, mesons scattered at large angles in the laboratory frame. The HARP experiment covered such a large-angle phase space with their central TPC.  Figure~\ref{fig:exampleOfCoverage} visualizes HARP's phase-space coverage in the $x_{F}$--$p_T$ plane. 
To discuss quantitatively, we used our MC code to evaluate the percentage of phase space not covered by any beam data. Beam data were grouped into six sections according to their beam momentum, as shown in Table~\ref{tab:beam_momentum_section}.  
\begin{table*}
  \caption{Sections of beam momentum, and experiments belonging to the section.}
 \label{tab:beam_momentum_section}
  \begin{center}
   \begin{tabular}{c | c | c : c : c : c : c : c }
    $p_{beam}$ [GeV/c] & experiment & \multicolumn{6}{c}{ section number }\\
     \hline
     3 & HARP~\cite{HARP,HARP_LA,HARP_proton} & \multirow{2}{*}{(I)} & & & & & \\
     \cline{1-2} \cline{4-4} \cdashline{5-8}
      5 & HARP~\cite{HARP,HARP_LA,HARP_proton} &  &\multirow{3}{*}{(II)} & & & & \\ 
      \cline{1-2} \cline{3-3}\cdashline{5-8}
      6.4 & E910~\cite{E910} & & & & & & \\
      \cline{1-2} \cdashline{3-3} \cline{5-5} \cdashline{6-8}
      8 & HARP~\cite{HARP,HARP_LA,HARP_proton} & & &\multirow{3}{*}{(III)} & & & \\
      \cline{1-2} \cdashline{3-3} \cline{4-4} \cline{6-6} \cdashline{7-8}
      12 & HARP~\cite{HARP,HARP_LA,HARP_proton} & & & &\multirow{3}{*}{(IV)}  & & \\
      \cline{1-2} \cdashline{3-4}\cdashline{7-8}
      12.3 & E910~\cite{E910} & & & & & & \\
      \cline{1-2} \cdashline{3-4}\cline{5-5}  \cline{7-7} \cdashline{8-8}
      17.5 & E910~\cite{E910} & & & & & \multirow{2}{*}{(V)} & \\
      \cline{1-2} \cdashline{3-5}\cline{6-6} \cline{8-8}
      31 & NA61/SHINE~\cite{NA61} & & & & & & \multirow{4}{*}{(VI)}\\
      \cline{1-2} \cdashline{3-6}\cline{7-7}
      158 & NA49~\cite{NA49} & & & & & & \\
      \cline{1-2} \cdashline{3-7}
      400 & NA20~\cite{BMPT} & & & & & & \\
      \cline{1-2} \cdashline{3-7}
      450 & NA56/SPY~\cite{BMPT} & & & & & & \\
      \cline{1-2} \cdashline{3-7}
    \hline 
    \end{tabular}
       \end{center}
       \end{table*}
Each secondary hadron was classified into one of these sections based on its parent particle momentum $p_{in}$, and we checked whether its $x_{F}$ and  $p_{T}$ were covered by at least one beam data belonging to that section. 
Figure~\ref{fig:coverage} shows a fraction of interactions whose phase space is not covered by any beam data. 
Nucleon-induced pion production, which occupies about half of all hadron interactions, is well covered with beam data in the $1<p_{\nu}<10$ GeV region. Phase-space coverage becomes worse in the $p_\nu < 1$ GeV region, owing to the large-angle scattering mentioned above. In addition, the fraction of nucleon production increases in this region, for which the beam data provide less phase-space coverage. The insufficient coverage of the phase space will be considered as the systematic uncertainty of the accelerator-driven tuning, as discussed in Section~\ref{sec:syst-uncert-1}. %Phase space coverage becomes worse in the $< 1$ GeV region owing to the large angle scattering mentioned above and the increase in nucleon production for which beam data provides less phase space. 

%The result of the phase space coverage evaluation is shown in Fig.~\ref{fig:coverage}. The coverage for pion production is relatively good (70--80\%) in multi-GeV $p_{\nu}$ region due to wide coverage of NA61 and NA49, but worse at lower momentum. Contributions of other particle productions to the uncoverage are larger than pion production. This phase space uncoverage was considered systematic uncertainty of this accelerator-driven tuning and will be discussed in Section~\ref{sec:syst-uncert-1}. 

\section{Parameterization}
\label{sec:parameterization}
In this section, we construct $[ E\frac{d^3\sigma}{dp^3}]_{data}$, the numerator on the right-hand side of Eq.~(\ref{eq:weight}) from the accelerator data listed in the previous section. 
The accelerator differential cross-section data have a finite binning, and the beam momenta are discrete. To be used as $[ E\frac{d^3\sigma}{dp^3}]_{data}$ in Eq.~\ref{eq:weight}, the data should be parameterized as a continuous function of $p_{in}$, $x_{F}$,  and $p_{T}$. %One of the difficulties of parameterization was that the interaction is a collision of nucleon--nucleus, and therefore we should consider the difference from nucleon--nucleon collision. 
We found that the entire dataset is difficult to describe with a single parameterization. Therefore, we divided the data into six sections based on their beam momentum, as shown in Table~\ref{tab:beam_momentum_section}. Parameterization was performed for each section separately, with a function defined below.

\subsection{Fitting function}
We determined the functional form of parameterization, $f_{p+A}$, as
\begin{equation}
 \label{eq:1}
  % f_{p+A}\left(p_{in},A, x_{F},p_{T}\right) \equiv f_{p+p}\left(x_{F},p_{T}\right)\times f_{nucl}\left(A,x_{F},p_{T}\right)  \times f_{inc}\left(p_{in}\right).
 f_{p+A}\left(p_{in},A, x_{F},p_{T}\right) \equiv f_{p+p}\left(|x_{F}|,p_{T}\right)\times f_{nucl}\left(A,x_{F},p_{T}\right)  \times f_{inc}\left(p_{in}\right).
\end{equation}
 $f_{p+p}$ describes the invariant differential cross-section of $p+p$ collision, which is symmetric in $x_{F}$ by construction. $f_{nucl}$ accounts for the asymmetry in $x_F$ arising from the fact that the target is an atomic nucleus rather than a nucleon, and it depends on the mass number $A$ of the target nucleus. $f_{inc}$ is added to describe the dependence on incident momentum. 
 
We used the BMPT parameterization introduced in~\cite{BMPT} to describe $f_{p+p}$ for $\pi$ and $K$ production. The parameterization is expressed as a function of $x_{R}$, which is called "radial scaling~\cite{RadialScaling}" and defined as $x_R \equiv E_{CM}/E_{CM}^{max}$. From the definition, $x_R$ is a function of absolute value of $x_F$: $x_R = x_R(|x_F|)$. The BMPT parameterization is 
\begin{eqnarray}
\label{eq:fpp_pik}
 f_{p+p}(|x_F|,p_T) &=& \nonumber \\
 f_{p+p}(x_R,p_T) &=& c_{1}\left(1-x_R\right)^{c_{2}}(1+c_{3}x_R){x_R}^{-c_{4}}\times \nonumber\\
  & & \left(1+D_1(x_R) p_T+D_2(x_R){p_T}^2\right)e^{-D_1(x_R) {p_T}},
\end{eqnarray}
where 
\begin{equation}
 D_1 = \frac{c_{5}}{{x_R}^{c_{6}}},\textrm{ and } D_2 = \frac{{c_{5}}^2}{2{x_R}^{c_{7}}}.
\end{equation}
This function has seven input parameters, $c_1$,..., $c_7$. Because it is a function of $|x_F|$, it is symmetric in $x_F$.
%The $x_R$ is {\it radial scaling} variable~\cite{}, defined as the ratio of the energy of a produced particle in the center-of-moment frame to its maximum vae kinematically available, {\it i.e.:} $x_R \equiv E_{CM}/E_{CM}^{max}$. The $x_R$ is one-to-one correspondence to $|x_F|$, thus $f_{p+p}$ has a symmetric form to $x_F$. 
For proton production, we use 
\begin{eqnarray}
\label{eq:3}
 f_{p+p}^{proton} &\equiv& c_{1}(1+c_{2}x_F+E_{1}(p_T)\times c_{3}x_F^2)\times \nonumber\\
 & & E_{2}(x_F) \times (1-x_F)^{c_{5}{p_T}^{c_{7}}}(1+c_{4}p_T+\frac{(c_{6}p_T)^2}{2})e^{-c_{4}p_T},
\end{eqnarray}
based on the function introduced in~\cite{BMPT}, with some modifications; we introduced $E_{1}(p_T)$ and $E_{2}(x_F)$ to explain the distribution observed in NA49. They are defined as %based on the function introduced in~\cite{BMPT} with adding some modification;
\begin{equation}
 E_{1} \equiv 
  \left\{ 
   \begin{array}{ll} 
    (1-p_T/c_8)^{c_9} & (p_T < c_8) \\
    0 & (\textrm{else})
   \end{array}
  \right.
\end{equation}
and
\begin{equation}
 E_{2}\equiv 1+c_{10} \exp\left(\frac{x_F-1}{c_{11}}\right).
\end{equation}
For data with beam momenta $< 31$ GeV/$c$, $E_1 = E_2 = 1$.

\begin{figure}[!th]
\centering
\includegraphics[width=0.75\textwidth]{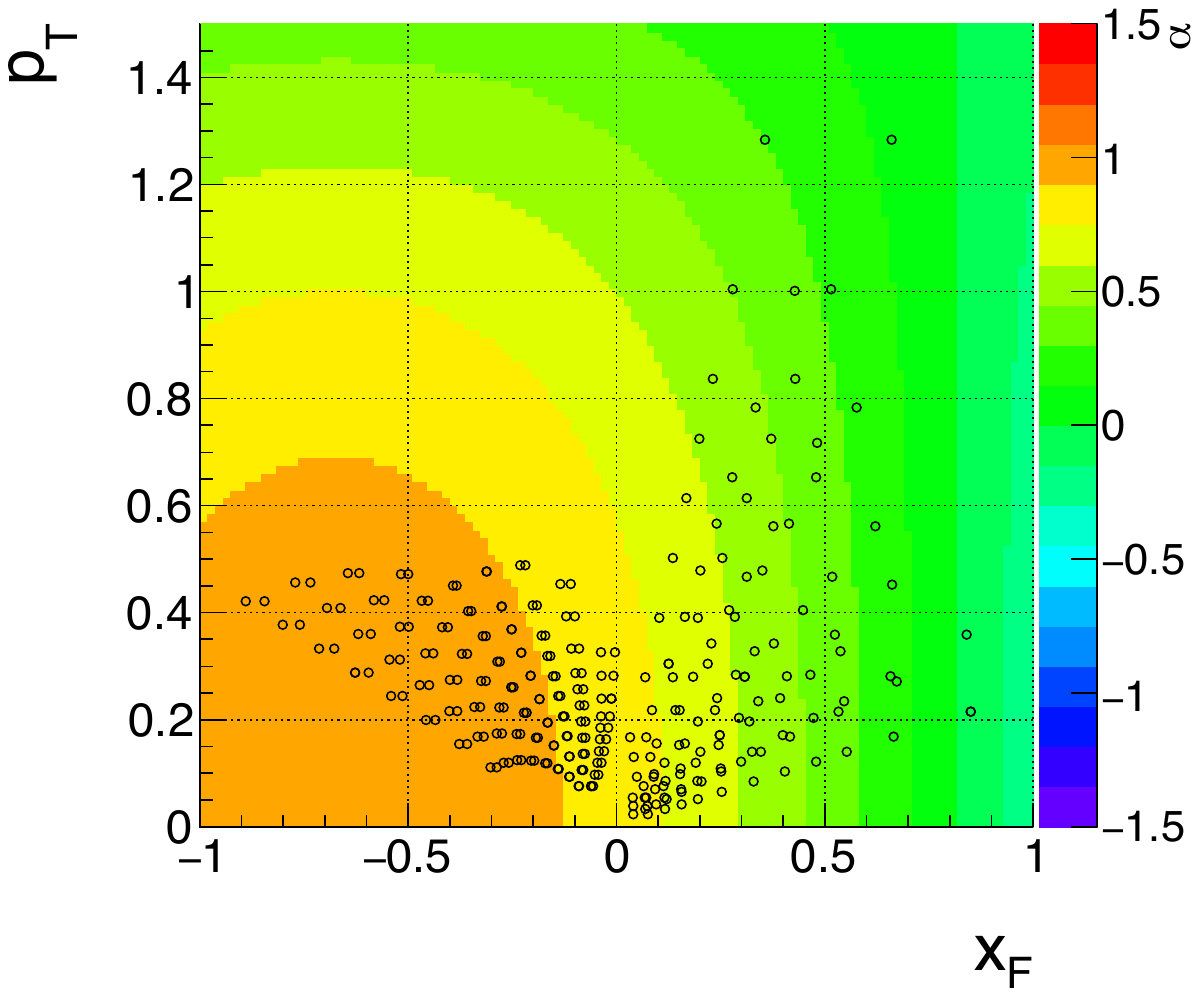}
\caption{$\alpha$ function for $\pi^+$ production in $p_{in} =$ 8--12 GeV/$c$ region. Black circles show data point.}
\label{fig:alphaFunc}
\end{figure}

We defined $f_{nucl}$ in Eq.~(\ref{eq:1}) as follows:
\begin{equation}
\label{eq:fnucl}
 f_{nucl} \equiv 
  \left\{ 
   \begin{array}{ll} 
    f_{p+A/p+d} \times f_{p+d/p+p} & (\textrm{$\pi$ or $K$ production}) \\
    f_{p+A/p+p}  & (\textrm{$p$ production})
   \end{array}
  \right.
\end{equation}
$f_{p+A/p+d}$ and $f_{p+A/p+p}$  describe the mass number dependence, defined as 
\begin{eqnarray}
\label{eq:14}
 f_{p+A/p+d} & = &  \left(\frac{A}{A_d}\right)^\alpha \\
 \label{eq:14_2}
 f_{p+A/p+p} & = &  \left(\frac{A}{A_p}\right)^\alpha, 
 \end{eqnarray}
where $A$, $A_d$, and $A_p$ are the mass numbers of the target nucleus, deuteron, and proton, respectively. This power-law form of mass number dependence does not extrapolate well to $A=1$ because of the isospin difference in the meson yield in proton-proton and proton-nucleon collisions. Thus, for the meson production, the power-law scaling is applied down to $A = A_d$ (deuteron) rather than to $A = 1$, and another function $f_{p+d}/f_{p+p}$ is multiplied to account for the isospin difference, as described later in Eq.~\ref{eq:pdpp}. For proton production, we simply used Eq.~\ref{eq:14_2}. 
$\alpha$ in Eqs.~\ref{eq:14} and \ref{eq:14_2} is parameterized as a function of $x_F$ and $p_T$~\cite{BMPT}: 
\begin{equation}
\label{eq:7}
\alpha \equiv (a_{1}+a_{2}x_F+a_{3}x_F^2)\times(1+a_{4}p_T^2).
\end{equation}
The parameters $a_1,...,a_4$ were determined using Be, C, and Al target data measured by the HARP experiment with beam momenta $p_{in} = $ 3, 5, 8, and 12 GeV/$c$. We calculated the ratio of differential cross-sections for the C (or Al) target to that for the Be target and fit the ratio with a function
\begin{equation}
 \left(\frac{A_{C}}{A_{Be}}\right)^\alpha \textrm{ or } \left(\frac{A_{Al}}{A_{Be}}\right)^\alpha,
\end{equation}
where $A_{Be}, A_{C}$, and $A_{Al}$ are mass numbers of Be, C, and Al, respectively. The fitting was separately conducted for each of $\pi^+$, $\pi^-$, and $p$ production data, and for each beam momentum section I, II, and III shown in Table~\ref{tab:beam_momentum_section}. The result of the fitting is summarized in Table~\ref{tab:alpha}. Figure~\ref{fig:alphaFunc} shows $\alpha$ for $\pi^+$ in the momentum section III ({\it i.e.}, $p_{in} = $ 8--12 GeV/$c$ region). For the momentum section IV-VI ({\it i.e.}, $p_{in} >$ 12 GeV/$c$ region), we used the parameter set for $\alpha$ derived from the 8--12 GeV/$c$ fitting. For $K^+$ and $K^-$ productions, we used the parameter sets for  8--12 GeV/c $\pi^+$ and $\pi^-$ productions, respectively, assuming the dependence of $\alpha$ on particle type is not large. For $x_F$--$p_T$ phase space not covered by HARP data, we extrapolated the $\alpha$ function. The uncertainties related to these substitutions and extrapolation will be evaluated as systematic uncertainties of the tuning in Section~\ref{sec:syst-uncert-1}.

\begin{figure}[!th]
\centering
\includegraphics[width=0.98\textwidth]{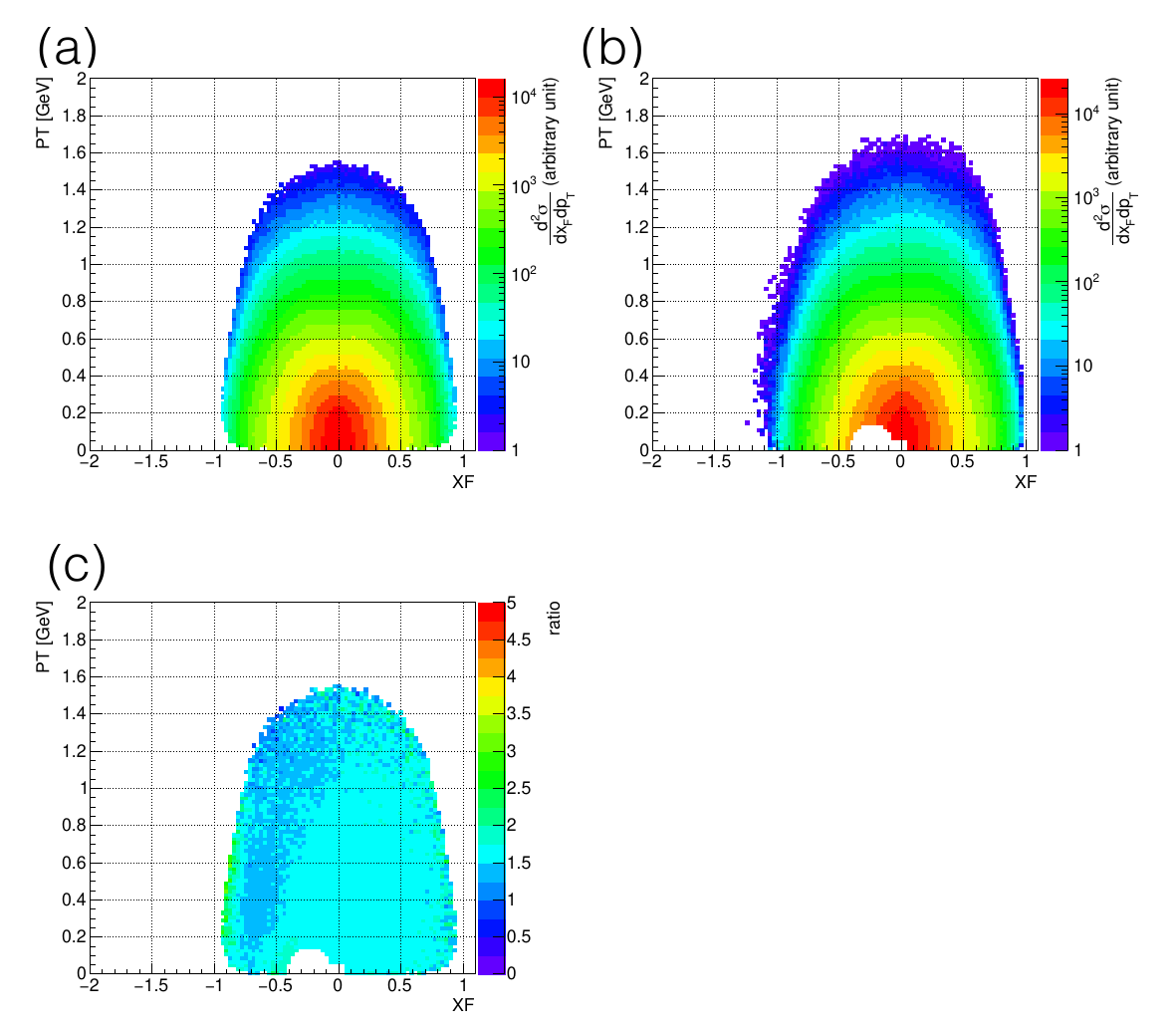}
\caption{MC expectations of the differential cross-section for $\pi^+$ production at $p_{in} =$ 12 GeV/$c$ for (a) $p+p$ collision and (b)  $p+d$ collision. The ratio of (b) to (a) is shown in (c). }
 \label{fig:fpapp}
\end{figure}

%In the case of meson production, we introduced another function $f_{p+d/p+p}$ in Eq.~\ref{eq:fnucl}, in order to explain the difference between the nucleus target ($A>1$) and proton target($A=1$). Since a nucleus target includes neutrons in contrast to proton target, the $f_{p+A/p+d}$ in Eq.~\ref{eq:14} doesn't extrapolate well to $A=1$. In the case of proton production $f_{p+d/p+p}$ is not introduced because the outgoing protons are mainly scattered protons. 
 As expressed in Eq.~\ref{eq:fnucl}, $f_{p+d/p+p}$ is introduced for the meson production. The functional form was empirically determined based on simulations for $p+p$ and $p+d$ collisions using DPMJET-III~\cite{DPM}. Figure~\ref{fig:fpapp} shows the simulated differential cross-sections of $p+p$ and $p+d$ collisions with incident proton momentum = 12 GeV/$c$ and its ratio. We determined the functional form of $f_{p+d/p+p}$ to roughly reproduce the difference between the $p+d$ and $p+p$ collisions as
\begin{equation}
\label{eq:pdpp}
f_{p+d/p+p}\equiv \exp\left(\sum_{i=0}^2\sum_{j=0}^2 b_{ij}x_{F}^ip_{T}^j\right),
\end{equation}
where $b_{00}$ is fixed to 0. We normalized $f_{p+d/p+p}$ to be unity at $x_F = 0.4$ and $p_T=0.05$ GeV/$c$. %For proton production we did not introduce $f_{p+d/p+p}$.  
%because it did not improve the  agreement with the data.

\begin{table*}
  \caption{Parameters of $\alpha$ function. To check consistency with data, $\chi^2$ with HARP measurements is also shown in the last column.}
\label{tab:alpha}
  \begin{center}
   \begin{tabular}{c |c | c | c | c | c | c |}
    PID & $p_{in}$ [GeV/$c$] & $a_1$ & $a_2$& $a_3$& $a_4$ &  $\chi^2$ \\
    \hline 
    $\pi^+$ & 3--5  & 0.670 & -0.927 & -0.1928 & -1.77392 & 0.868\\
            & 5--8  & 0.772 & -0.660 & 0.00802 &  -0.897 & 0.973\\
            & 8--12 & 0.825 & -0.623 & -0.463  &  -0.276 & 0.823\\
            & $>12$ & \multicolumn{4}{c|}{same as 8-12 GeV/c}& -\\
    \hline 
    $\pi^-$ & 3--5  & 0.689 & -0.617 & -1.12 & -0.957 & 1.20\\
            & 5--8  & 0.792 & -0.529 & -1.04 & -0.697 & 1.13\\
            & 8--12 & 0.854 & -0.623 & -1.22 & -0.629 & 0.822\\
            & $>12$ & \multicolumn{4}{c|}{same as 8-12 GeV/c}& -\\
    \hline 
    $K^+$ & 3--5  & \multicolumn{4}{c|}{\multirow{4}{*}{same as $\pi^+$}}&-\\
            & 5--8  & \multicolumn{4}{c|}{}&-\\
            & 8--12 & \multicolumn{4}{c|}{}&-\\
            & $>12$ & \multicolumn{4}{c|}{}&-\\
    \hline 
    $K^-$ & 3--5  & \multicolumn{4}{c|}{\multirow{4}{*}{same as $\pi^-$}}&-\\
            & 5--8  & \multicolumn{4}{c|}{}&-\\
            & 8--12 & \multicolumn{4}{c|}{}&-\\
            & $>12$ & \multicolumn{4}{c|}{}&-\\
    \hline 
    $p$     & 3--5  & 0.653 & -0.773 & 0.502 & 0.731 & 0.504\\
            & 5--8  & 0.633 & -0.651 & 0.391 & 0.536 & 0.562\\
            & 8--12 & 0.677 & -0.610 & 0.255 & 0.458 & 0.619\\
            & $>12$ & \multicolumn{4}{c|}{same as 8-12 GeV/c}& -\\
    \hline
       \end{tabular}
  \end{center}
 \end{table*}

$f_{inc}$ in Eq.~(\ref{eq:1}) is a function to describe energy dependence. Because we divided the data into small momentum sections, a simple logarithmic interpolation describes the data well in the section. $f_{inc}$ is defined as
\begin{equation}
\label{eq:10}
f_{inc} \equiv k \log_{10}\frac{p_{in}}{p_0} + 1,
\end{equation}
where $p_0$ is the maximum beam momentum in the fitted data, and $k$ is a free parameter. 

\subsection{Fitting}

\begin{figure}[!th]
\centering
\includegraphics[width=0.98\textwidth]{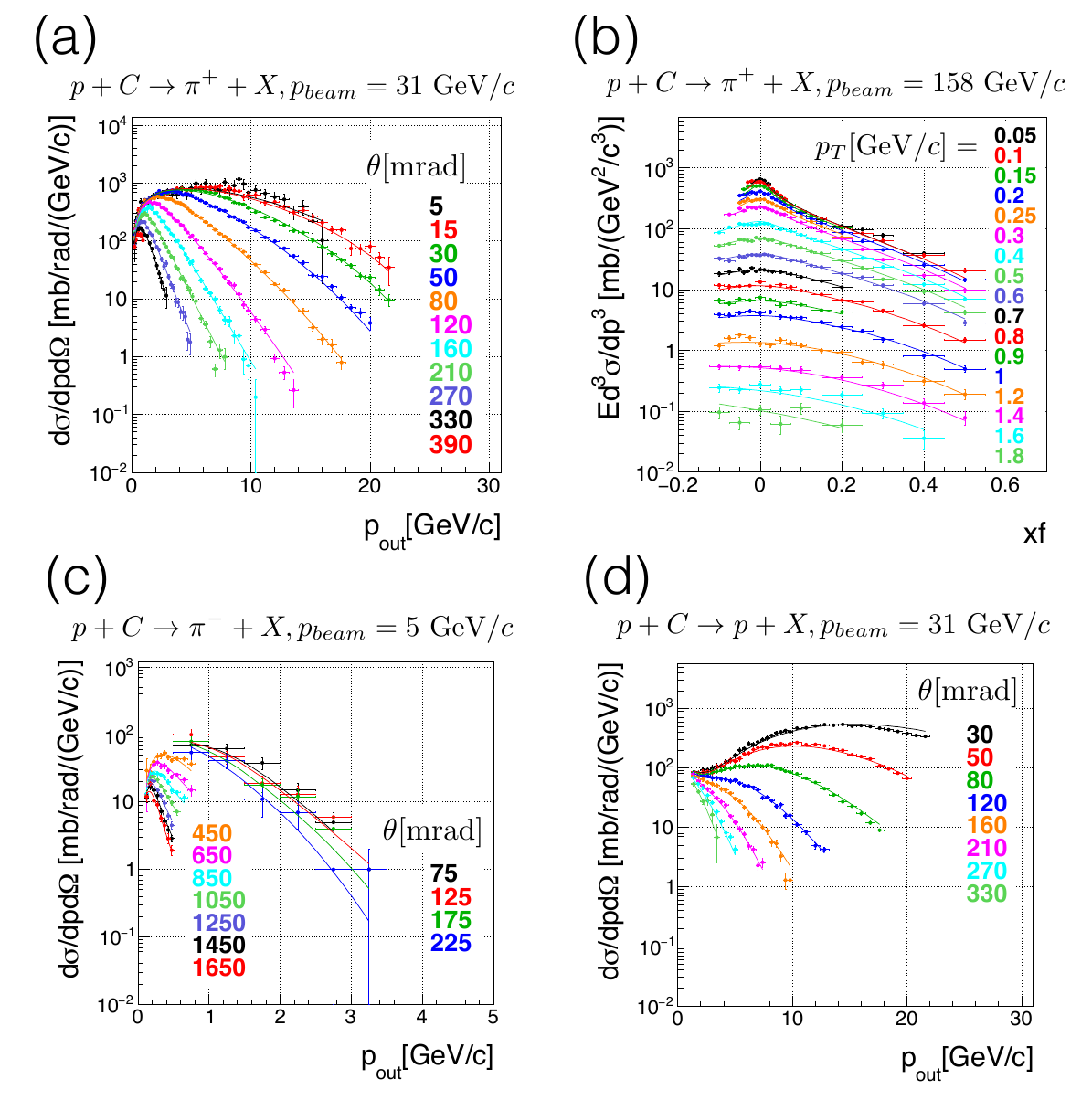}
\caption{Examples of fit results. Each example is taken from the fit results for (a) $\pi^+$ production at 31 GeV/$c$ section, (b) $\pi^+$ production at 158 GeV/$c$, (c) $\pi^-$ production at 5 GeV/$c$, (d) proton production at 31 GeV/$c$. The points with error bars show data, while solid lines show the $f_{p+A}$ function. The color differences represent different projectile angles (projectile $p_{T}$ for (b)) given inside each plot.}
\label{fig:fitResult}
\end{figure}

We fitted the data with $f_{p+A}$ in Eq.~(\ref{eq:1}) for each of the six momentum sections separately.
For each section, we minimized $\chi^2$, which is defined as
\begin{equation}
\label{eq:9}
% \chi^2 \equiv \sum_{i}^{data}\frac{\left( \frac{d^2\sigma}{dpd\Omega}_i - <\frac{p_{out}^{2}}{E_{out}}f_{IDCS}>_i \right)^2}{\sigma_i^2} + \sum_{j}^{beam}\left(\frac{\Delta_j^2}{\sigma_j^2}\right),
 \chi^2 \equiv \sum_{j}^{beam}\sum_{i}^{bin}\frac{\left( (1+\Delta_j)y_{ji} - \langle f_{p+A} \rangle_{ji} \right)^2}{\sigma_{ji}^2} + \sum_{j}^{beam}\left(\frac{\Delta_j^2}{\sigma_{\Delta j}^2}\right),
\end{equation}
where $y^{data}_{ji}$ and $\sigma_{ji}$ are the measured differential cross-section of the $i$-th bin in the $j$-th beam data and its measurement error respectively, and $\langle f_{p+A}\rangle_{ji} $ is an averaged value of $f_{p+A}$ inside the $i$-th bin in the $j$-th beam data. The summation $\sum_{j}^{beam}$ runs over the beam datasets belonging to the momentum section being fitted. $\sum_{j}^{beam}\left(\Delta_j^2 /\sigma_{\Delta j}^2 \right)$ accounts for the normalization uncertainty of the data. In the fitting process, parameters $a_1,...,a_4$ of $f_{p+A/p+d}$ in Eq.~(\ref{eq:7}) were fixed as shown in Table~\ref{tab:alpha}. $p_0$ in Eq.~(\ref{eq:10}) was fixed to the largest beam momentum in the fitting section. For $\pi$ or $K$ production, free parameters are $\{c_1,...,c_7\}$ in Eq.~(\ref{eq:fpp_pik}), $\{b_{01},...,b_{22}\}$ in Eq.~(\ref{eq:pdpp}), and $k$ in Eq.~(\ref{eq:10}). For $p$ production, the free parameters are $\{c_1,...,c_{11}\}$ in Eq.~(\ref{eq:3}) and $k$ in Eq.~(\ref{eq:10}). %explains the uncertainty of normalization of the data
Some examples of fitting results  are shown in Fig.~\ref{fig:fitResult}.
We successfully fitted the data with reduced $\chi^2$ values around 2 or less in all the sections, as summarized in Table~\ref{tab:chi2}. 

The fitted function $f_{p+A}$ obtained here is evaluated at discrete momenta $p_{w}$ to construct $[ E\frac{d^3\sigma}{dp^3}]_{data}$ in Eq.~(\ref{eq:weight}), as described in the next section.

\begin{table*}
  \caption{Reduced $\chi^2$ of fitting in each beam momentum section for each production particle.}
 \label{tab:chi2}
  \begin{center}
   \begin{tabular}{c | c | c | c | c | c | c |}
       &  \multicolumn{6}{c|}{beam momentum section [GeV/c]} \\
  & (1) & (2) & (3) & (4) & (5) & (6)\\ 
   & 3--5 & 5--8 & 8--12.3 & 12--17.5 & 17.5--31 & $>$ 31\\ 
    \hline 
    \hline 
    $\pi^{+}$  & 1.43 & 1.63 & 1.72 & 1.80 & 1.96 & 1.79\\
       & (549)&(732)&(889)&(554)&(471)&(691) \\
    \hline 
    $\pi^{-}$  & 1.41 & 1.53 & 1.51 & 1.57 & 1.25 & 2.10\\
         & (510)&(683)&(838)&(528)&(504)&(716) \\
    \hline 
    $K^{+}$  & -- & -- & -- & -- & -- & 0.80\\
       &    &    &    &    &    &(103) \\
    \hline 
    $K^{-}$  & -- & -- & -- & -- & -- & 1.34\\
        &    &    &    &    &    &(89) \\
    \hline 
    $p$  & 1.02 & 1.66 & 1.50 & \multicolumn{2}{c|}{2.24} & 1.26\\
     &(119) &(179) &(215) &\multicolumn{2}{c|}{(293)}&(200) \\
    \hline 

       \end{tabular}
  \end{center}
 \end{table*}

\section{Derivation of the weight function}
\label{sec:accel-data-driv-2}
%\subsection{Derivation of the weight}
In the previous section, we obtained $f_{p+A}$ from the fitting, which is used to derive the $[ E\frac{d^3\sigma}{dp^3}]_{data}$, the numerator on the right-hand side of Eq.~\ref{eq:weight}. In this section, we construct the denominator $[\sigma_{prod}]_{MC}$ and $ [E\frac{d^3n}{dp^3}]_{MC}$ and, consequently, the weight function $w$ in Eq.~\ref{eq:weight}. 
The weight function $w$ is separately constructed for each combination of incident particle $x_{in}$ and secondary particle $x_{out}$ listed in Table~\ref{tab:combination}. 

  $[E\frac{d^3 n}{dp^3}]_{MC}$ was evaluated by running our MC code. We generated $N_{gen} = 4.8\times 10^{8}$ hadron interactions for each particle combination listed in Table~\ref{tab:combination} and for each of the following incident momenta: $p_{w} =$ every 0.5 GeV/$c$ between 3--10 GeV/$c$ and 11, 12, 14, 16, 17.5, 20, 25, 31, 33, 50, 75, 100, 158, 200, 300, and 450 GeV/$c$. 
  For $[\sigma_{prod}]_{MC}$, we used the hadron production cross-section implemented in our MC simulation.

 \begin{table*}
 \caption{Combination of incident particle $x_{in}$ and secondary particle $x_{out}$. "Data" means the beam experiments provide the data. "ISO" and "QPM" indicate differential cross-sections derived using isospin relation and quark counting, respectively. }
 \label{tab:combination}
  \begin{center}
   \begin{tabular}{c | c | c | c | c |c |c|c}

    \diagbox{$x_{in}$}{$x_{out}$} & $\pi^+$ & $\pi^-$ & $K^+$ & $K^-$ & $K_{S}, K_{L}$ & p & n \\
    \hline
    p & data & data & data & data & QPM & data & assumption \\
    \hline
    n & ISO & ISO & QPM & QPM & ISO & ISO & assumption \\
   \end{tabular}
  \end{center}
 \end{table*}
 
To derive $[E\frac{d^3 \sigma}{dp^3}]_{data}$, the fitted function $f_{p+A}$ was evaluated at each discrete momentum $p_{w}$. If $p_w$ lies at the boundary between two momentum sections, the $f_{p+A}$ values from the two adjacent sections are averaged (for example, for $p_w = 5$ GeV/$c$, $f_{p+A}$ values for section (1) 3-5 GeV/$c$ and section (2) 5-8 GeV/$c$ are averaged). The particle combinations listed in Table~\ref{tab:combination} include some for which we do not have beam data. For such a combination, we use isospin relations: %If $p_w$ is a boundary of the momentum sections, an average of two sections is taken
\begin{eqnarray}
    \sigma(n&+&A_{air} \to \pi^{\pm} + X) = \sigma(p+A_{air} \to \pi^{\mp} + X) \\
    \sigma(n&+&A_{air} \to n + X) =   \sigma(p+A_{air} \to p + X)\\
    \sigma(n&+&A_{air} \to K_{S,L} + X)\nonumber \\ 
    &=& \frac{1}{2}\{ \sigma(p+A_{air} \to K^{+} + X) + \sigma(p+A_{air} \to K^{-} + X) \},
\end{eqnarray}
and the quark counting method based on simple quark parton model: 
\begin{eqnarray}
    \sigma(p&+&A_{air} \to K_{S,L} + X)  \nonumber \\ 
    &=& \frac{1}{8}\{ 3\sigma(p+A_{air} \to K^{+} + X) + 5\sigma(p+A_{air} \to K^{-} + X) \} \\
     \sigma(n&+&A_{air} \to K^{-} + X) \nonumber \\ &=&\sigma(p+A_{air} \to K^{-} + X) \\
     \sigma(n&+&A_{air} \to K^{+} + X) \nonumber \\
     &=& \frac{1}{4}\{ 3\sigma(p+A_{air} \to K^{+} + X) + \sigma(p+A_{air} \to K^{-} + X) \}.
\end{eqnarray}

Thus, all variables on the right-hand side of Eq.~(\ref{eq:weight}) are available, and we obtain $w$ for each $p_{w}$ and for each particle combination. The weight for a given incident momentum $p_{in}$ was then calculated with interpolation, as follows: 
\begin{equation}
 w(p_{in},x_F,p_T) = (1-F)\times w(p_{w_1},x_F,p_T)+ F\times w(p_{w_2},x_F,p_T),
\end{equation}
where $p_{w_1}$ and $p_{w_2}$ are the two $p_{w}$ values closest to a given $p_{in}$, and 
\begin{equation}
 F = \frac{\log p_{in} - \log p_{w_1}}{\log p_{w_2} - \log p_{w_1}}.
\end{equation}
%\begin{eqnarray}
%w_{x_{in}\to x_{out}}(p_{in},x_F,p_T) &=& \frac{w(p_{MC2},x_F,p_T)-w(p_{MC1},x_F,p_T)}{ \nonumber\\ &&+w(p_{MC1},x_F,p_T),
%\end{eqnarray}
%where $p_{MC1}$ and $p_{MC2}$ are the two $p_{MC}$'s closest to a given $p_{in}$. 

%For the interaction with $x_{in}$ = neutron for which we have no beam data, we calculated DCS from that of $x_{in}$ = proton.

\begin{figure}[!th]
\centering
\includegraphics[width=0.98\textwidth]{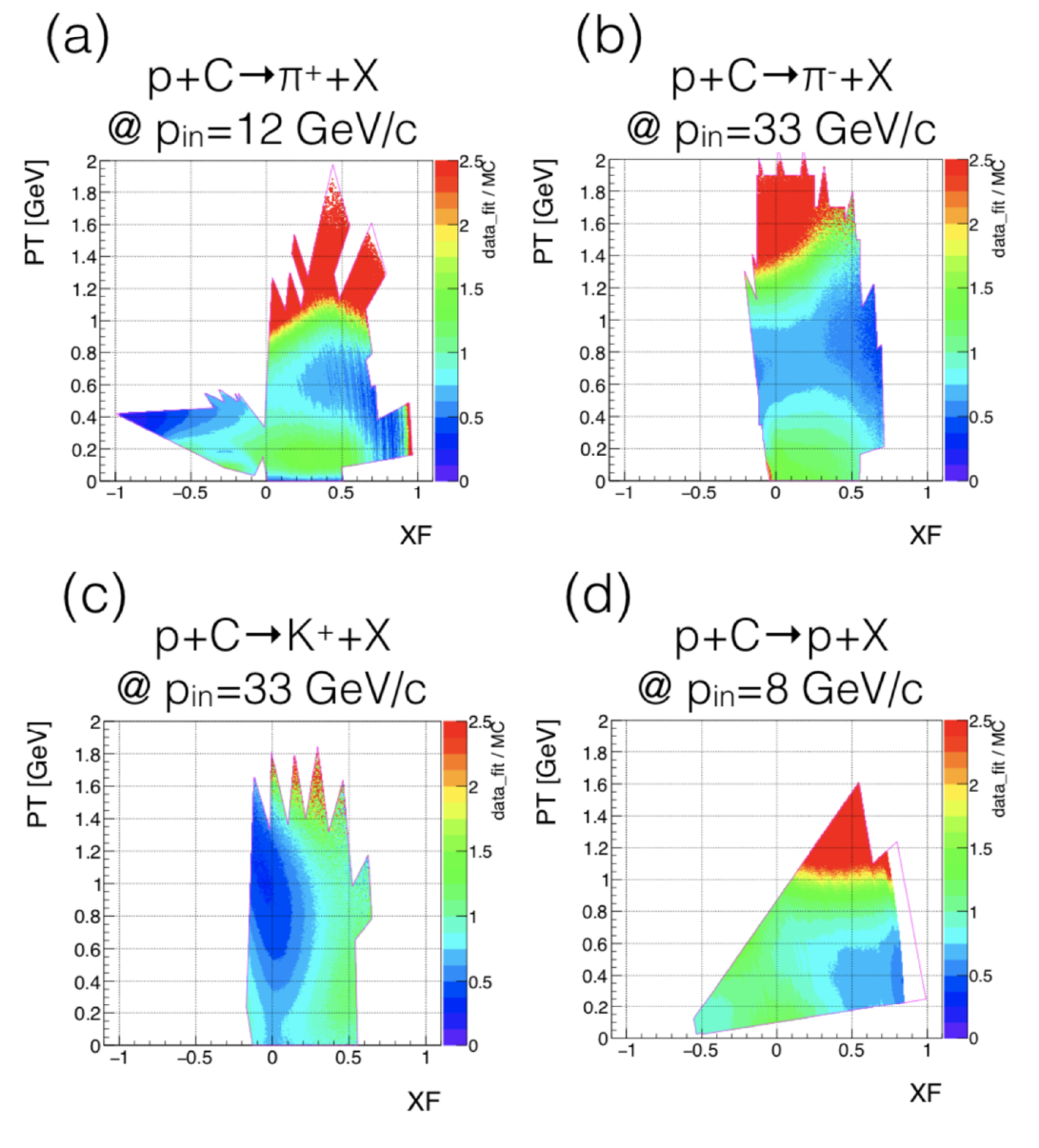}
\caption{Weight $w(p_{in},x_{in},x_{out}; x_F,p_T)$ in Eq.~(\ref{eq:weight}) as a function of $x_F$ and $p_T$. The uncolored region is not covered by any beam data, where the weight is set to 1. Four plots (a)-(d) show different $p_{in}, x_{in},$ and $x_{out}$. $(p_{in},x_{in},x_{out}) = $  $(12\textrm{ GeV/c}, p , \pi^+)$ in (a), $(33\textrm{ GeV/c}, p , \pi^-)$ in (b), $(33\textrm{ GeV/c}, p , K^+)$ in (c), and $(8\textrm{ GeV/c}, p , \pi^+)$ in (d).}
\label{fig:weightTable}
\end{figure}

For interactions with $450 <p_{in} < 1000$ GeV/$c$, we use $w$ for $p_{in} = 450$ GeV/$c$ under the assumption that Feynman scaling is well satisfied in such high-energy interactions. For $p_{in} < 3$ GeV/$c$ for pion or proton production, or for $p_{in} < 31$ GeV/$c$ for kaon production, we set $w = 1$. We set $w = 1$ if no beam data used for the fitting covers the given ($x_F, p_T$). Some examples of the weights are shown in Fig.~\ref{fig:weightTable}, where the uncolored region represents the phase space that no beam data covers. % For a given ($x_F,p_T$), we required that at least one of the beam data used for the fitting covers the ($x_F,p_T$), otherwise $w = 1$. %with assuming that Feynman scaling is well conserved 
For interactions $p +A_{air}\to n+X$ and $n +A_{air}\to p+X$, we assumed that the weight for  $p +A_{air}\to p + X$ can be applied. 

\subsection{Tuning the neutrino flux with the weight}
\begin{figure}[!th]
\centering
\subfigure[$\nu_\mu$ flux.]{\includegraphics[width=0.45\textwidth]{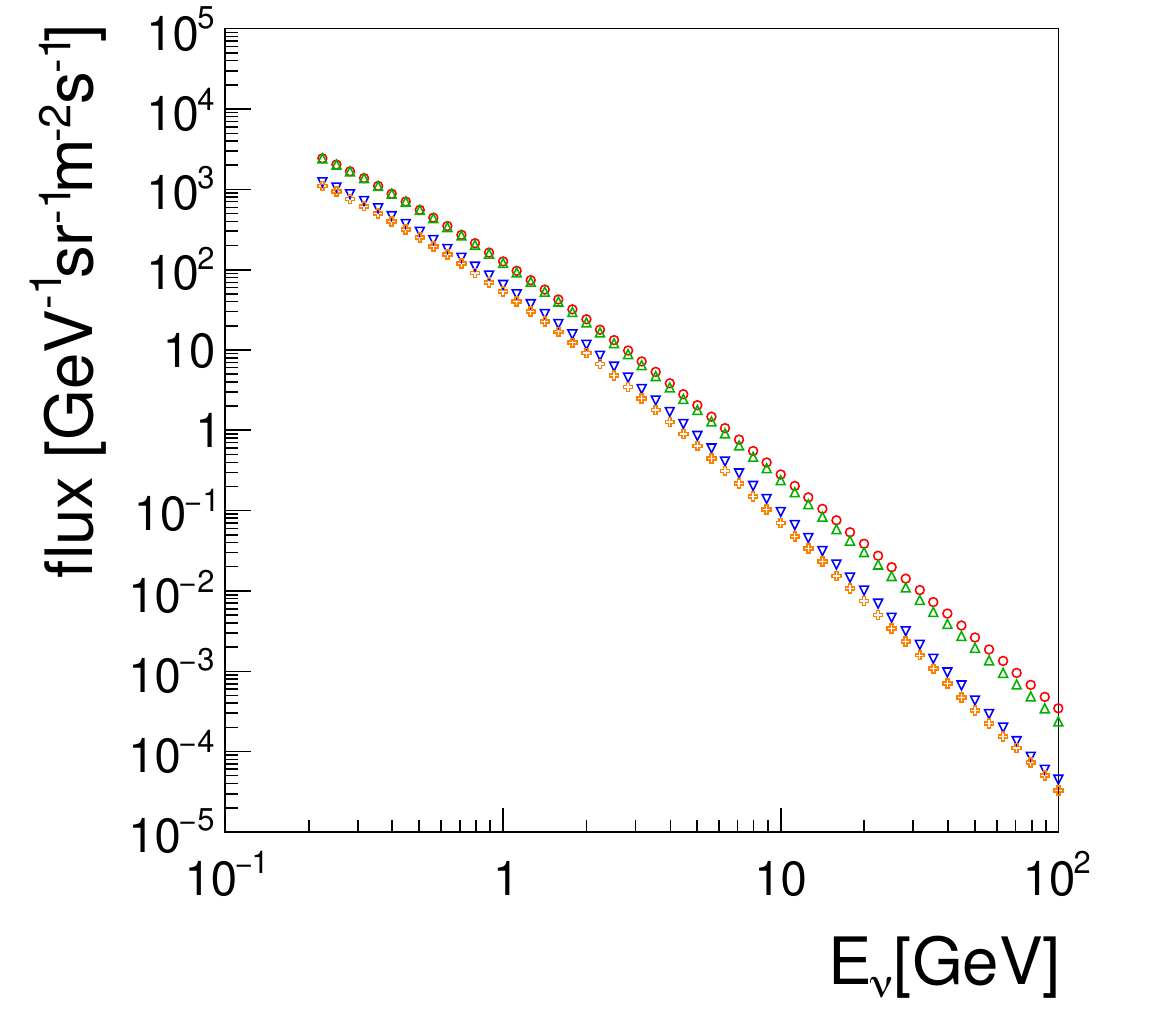}}
\subfigure[Flux ratio to the previous flux prediction]{\includegraphics[width=0.45\textwidth]{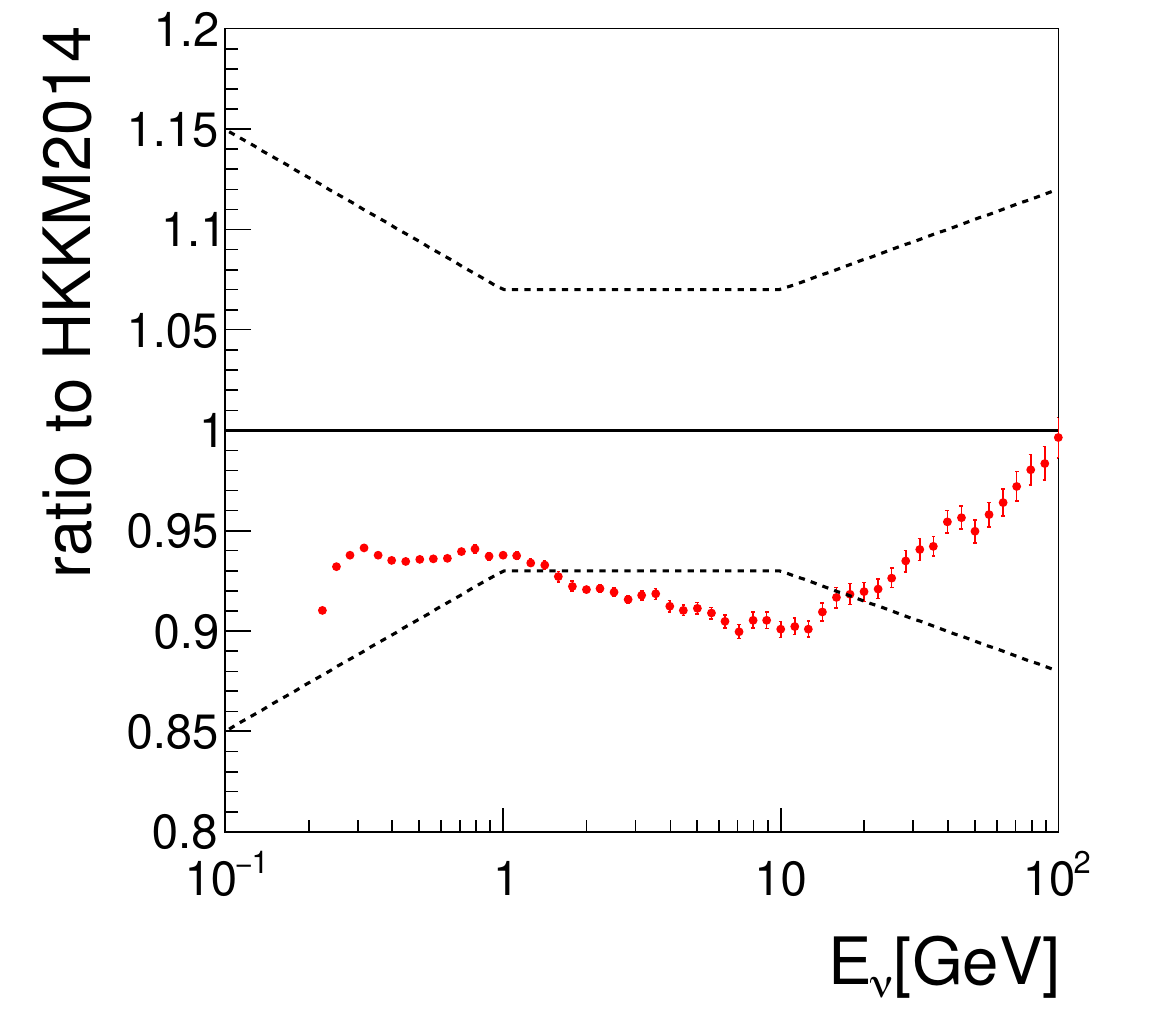}}
\caption{(a) Flux predictions with accelerator tuning. The red, green, blue, and orange lines correspond to $\nu_{\mu}$, $\bar{\nu}_{\mu}$, $\nu_{e}$, and $\bar{\nu}_{e}$, respectively. The error bars show the MC statistical error. (b) Ratio to the ``$\mu$-tuned'' flux prediction~\cite{HKKM2015}. The dashed line shows the systematic uncertainty reported in~\cite{HKKM2006}.}
\label{fig:flux}
\end{figure}

\begin{figure}[!th]
\centering
\includegraphics[width=0.98\textwidth]{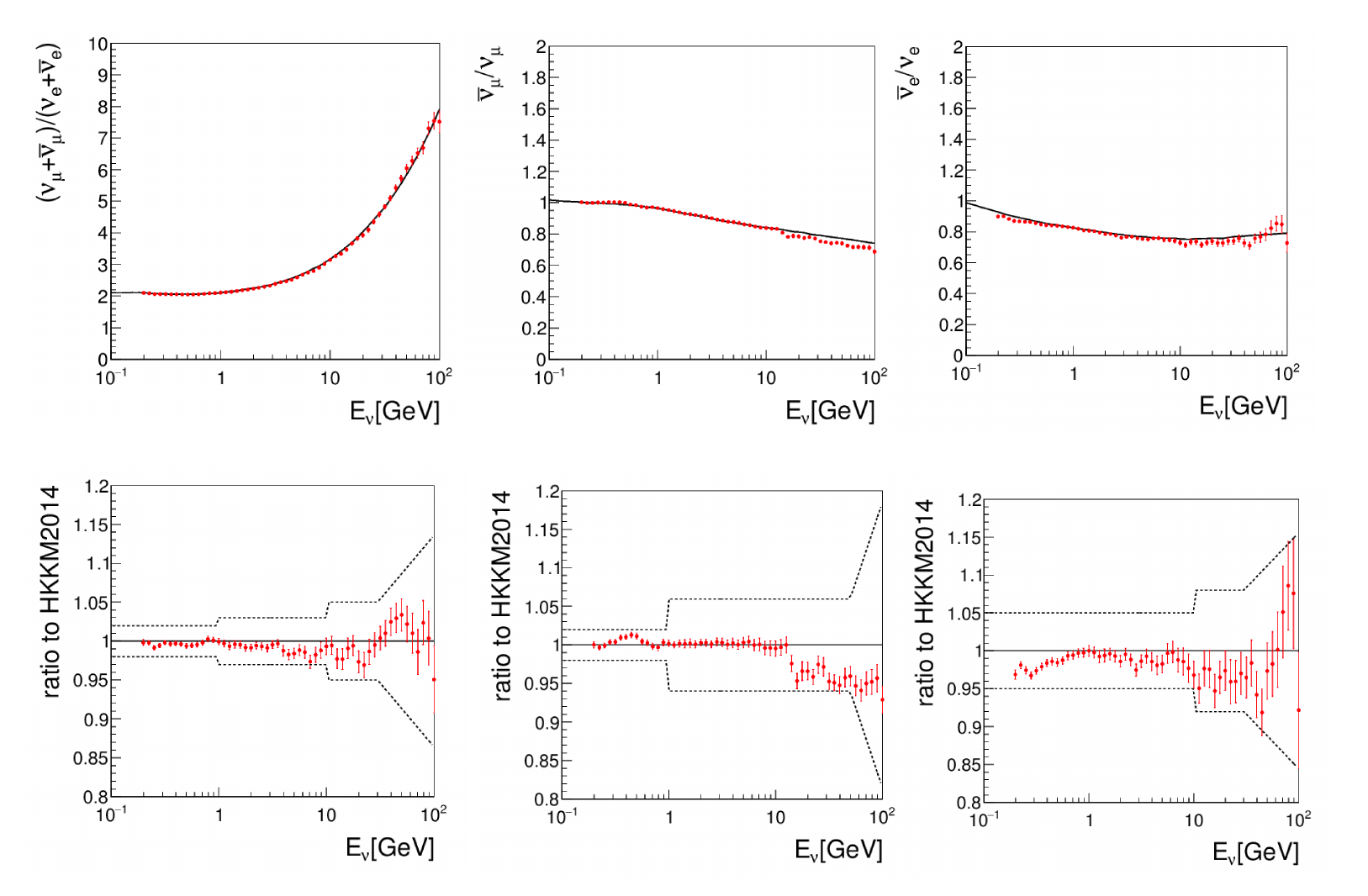}
 \caption{Flux ratio predictions with accelerator tuning. On the top panel, the flavor ratio (left), $\bar{\nu}_{\mu}/\nu_{\mu}$ ratio (middle), and $\bar{\nu}_{e}/\nu_{e}$ ratio (right) are shown.  The red dots show our accelerator tuning predictions, while the black line is the ``$\mu$-tuned'' predictions~\cite{HKKM2015}. On the bottom panel, the ratio of the accelerator tuning to the muon tuning is shown. The dashed lines are uncertainties of flux ratios used in Super-Kamiokande analysis.}
\label{fig:fluxRatio}
\end{figure}

We simulated the neutrino flux by applying the weight in Eq.~(\ref{eq:weight}). The result is shown in Fig.~\ref{fig:flux}. The flux is smaller by $\sim$5\%--10\% than the one previously reported in Ref.~\cite{HKKM2015}. The difference gradually disappears as the energy increases above 10 GeV. This is owing to the increasing contribution of interactions with $p_{in}>1000$ GeV, which is outside the scope of the accelerator data, so that the weight is set to be unity. This difference will be discussed later after evaluating the flux error in the next section.  %The flux is almost consistent with the one previously reported in Ref.~\cite{HKKM2015} considering its systematic error, though it has a tendency to be $\sim$5--10\% smaller as also shown in the figure. 
The $\nu_{\mu} / \nu_{e}$, $\bar{\nu}_{\mu} / \nu_{\mu}$, and $\bar{\nu}_{e} / \nu_{e}$ ratios are also shown in Fig.~\ref{fig:fluxRatio}. They are consistent with Ref.~\cite{HKKM2015} within a few \%.

% \begin{figure}[t]
% \centering
% \includegraphics[width=0.98\textwidth]{img/fluxSummary2.pdf}
%  \caption{(a) Flux predictions with accelerator tuning. The red, green, blue, and orenge correcspond to $\nu_{\mu}$, $\bar{\nu}_{\mu}$, $\nu_{e}$, and $\bar{\nu}_{e}$, respectively. Thier error bars shows MC statistical error only. (b) Ratio to the ``$\mu$-tuned'' flux prediction~\cite{HKKM2015}. Dashed line shows the systematic uncertainty reported in~\cite{HKKM2006}.}
% \label{fig:flux}
% \end{figure}

\section{Systematic uncertainty}
\label{sec:syst-uncert-1}

Several uncertainty sources are considered, as listed in Table~\ref{tab:Uncertainty}. The $I$-th source can modify the weight as
\begin{equation}
\label{eq:11}
 w\to w'_I = w\times (1 \pm \varepsilon_I).
\end{equation}
The modification $\varepsilon_I$ was evaluated separately for each uncertainty source and each $x_{in}$ and $x_{out}$ combination. We evaluated $\varepsilon_I$ based on experimental data where possible. Otherwise, we used a DPMJET 3.06~\cite{DPM} implemented in CRMC v1.6.0~\cite{CRMC} for evaluation. $\varepsilon_I$ is a function of $\left(p_{in},x_{F},p_{T}\right)$, except uncertainties related to normalization (Sys. 3 and 6), which are functions depending on $p_{in}$ only. This is shown in the ``type'' column in Table~\ref{tab:Uncertainty}. The ``sided'' column in the table indicates whether the $\varepsilon_{I}$ is double-sided or single-sided. If the ``sided'' column is ``$\pm$'' (double sided), modifications with both $+\varepsilon_{I}$ and $-\varepsilon_I$ are considered, whereas if it is "$+$" (single sided) only $+\varepsilon_{I}$ is considered.  
After $\varepsilon_I$ was evaluated for each uncertainty, we re-simulated the flux using the modified weight $w'_{I}$. The difference from the flux with default weight is considered a systematic uncertainty related to the $I$-th source. %

%\begin{landscape}
\begin{table*}
 \caption{Sources of flux uncertainties. See the text for the meaning of ``sided'' and ``type'' columns.}
 \label{tab:Uncertainty}
  \begin{center}
   \begin{tabular}{c | c | c | c | c }

    serial $I$& name & mom. range & sided & type \\
    \hline
    \hline 
    Sys. 1 & measurement normalization & all & $\pm$ & $p_{in}$ \\% & data \\
    Sys. 2 & measurement error in bins & all & $\pm$ & $p_{in},x_{F},p_{T}$ \\% & data \\
    Sys. 3 & phase space coverage  & 3--1000 GeV/c & + & $p_{in},x_{F},p_{T}$ \\% & CRMC \\
    Sys. 4 & phase space coverage  & $<3$ GeV/c, $>1000$ GeV/c & + & $p_{in},x_{F},p_{T}$ \\% & CRMC \\
    Sys. 5 & fitting $\chi^2$& all & + & $p_{in},x_{F},p_{T}$ \\% & data \\
%    interpolation in fitting & all & Yes & function & data \\
    Sys. 6 & Feynman scaling ($x_{F}-p_{T}$)& $>$ 158 GeV/c & + & $p_{in},x_{F},p_{T}$ \\% & CRMC \\
    Sys. 7 & Feynman scaling (normalization)& $>$ 400 GeV/c & + & $p_{in}$ \\% & CRMC \\
    Sys. 8 & $\alpha$ fitting error& all & $\pm$ & $p_{in},x_{F},p_{T}$ \\% & data \\
    Sys. 9 & $\alpha$ function for out-of-range & all & + & $p_{in},x_{F},p_{T}$ \\% & CRMC \\
    Sys. 10 & $\alpha$ function extrapolation & all & + & $p_{in},x_{F},p_{T}$ \\% & CRMC \\
   \end{tabular}
  \end{center}
 \end{table*}
%\end{landscape}

Each systematic uncertainty source is explained below. 

\begin{itemize}
 \item Measurement error of beam data (Sys. 1 and 2 in Table~\ref{tab:Uncertainty})

The measurement errors of differential cross-section are reported in each beam data paper. They are considered as systematic uncertainty sources of our flux study. We considered an uncertainty of overall normalization factor and errors in bins independently. 
The uncertainty of the overall normalization factor reported in measurement papers is shown in Fig.~\ref{fig:sys_measErr}(a) as a function of $p_{in}$, where gaps between the discrete beam momenta are interpolated logarithmically. This function is used as $\varepsilon_{1}$ in Eq.~(\ref{eq:11}). 
For errors in bins, measurement errors of $p+C$ data provided from HARP, NA61, and NA49  were used for evaluation.  The errors relative to differential cross-sections are used as $\varepsilon_{2}$ in Eq.~(\ref{eq:11}). An example of the relative error is shown as a 2-D histogram in the $x_F$ vs.\ $p_T$ plane in Fig.~\ref{fig:sys_measErr}(b).

\begin{figure}[!th]
\centering
\includegraphics[width=0.98\textwidth]{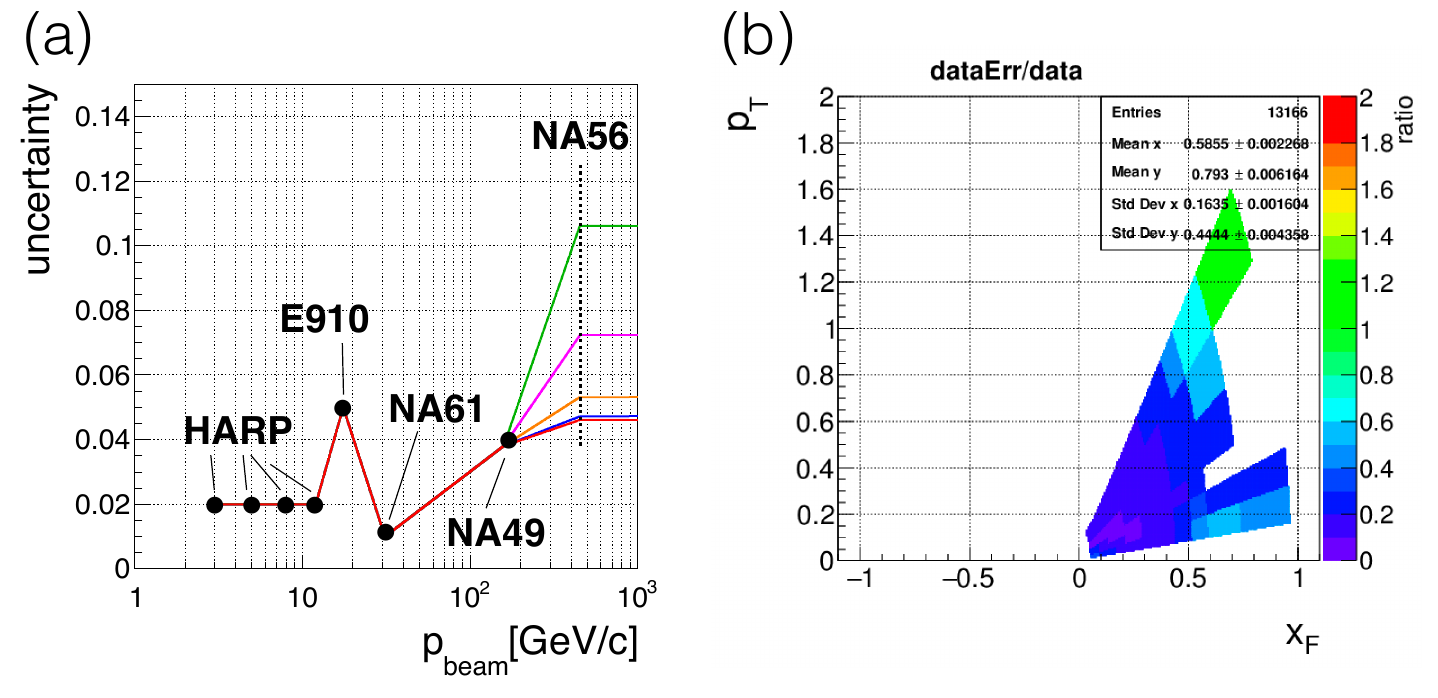}
 \caption{(a) Overall normalization uncertainty in measurements. Below 158 GeV/c, the same value is used for all particles. Above 158 GeV/c, red, green, blue, orange, and magenta represent uncertainties for $\pi^+, \pi^-, K^+, K^-$, and $p$ productions, respectively. (b) Relative error of differential cross-section measured in HARP $p+C$ 5 GeV data. The z-axis shows the measurement error divided by the measured value of differential cross-sections.}
\label{fig:sys_measErr}
\end{figure}

\item Phase space coverage (Sys. 3 and 4).
 
As discussed in Section~\ref{sec:accelerator-data}, accelerator data do not cover the entire phase space relevant to neutrino production. In such an uncovered phase-space, the weight was not applied. To evaluate the uncertainty due to the shortage of coverage, we expanded the weighting region. In $3 \le p_{in} \le 1000$ GeV/$c$ region, if the data cover a region $x_{min} < x_{F} < x_{max}$ at a given $p_{T}$ and $p_{in}$, the weight at $x_F = x_{min}$ was used for $x_F < x_{min}$, and the weight at $x_F = x_{max}$ was used for $x_F > x_{max}$. For $p_{in} < 3$ GeV/$c$ or $> 1000$ GeV/$c$, where no data were available, we used the weight for $p_{in} = 3$ GeV/$c$ or 1000 GeV/$c$ for the uncertainty evaluation. The difference of the fluxes applied these expanded weights from the flux with default weight was considered as systematic uncertainty. %the weight at $(x_{min},p_{T},p_{in})$ and $(x_{max},p_{T},p_{in})$ was used for $x_{F} < x_{min}$ and $x_{F} > x_{max}$ regions, respectively.

\item Fitting $\chi^2$ (Sys. 5) %Incompleteness of fitting function

In the fitting process described in Section~\ref{sec:parameterization}, the reduced $\chi^2$ values of the fitting were approximately 2 for almost all sections. The deviation of the reduced $\chi^2$ from unity suggests that the functional form used in the fitting does not perfectly describe the data, and this should be a source of systematic uncertainty. 
The simplest method to bring the reduced $\chi^2$ to unity is modifying $f_{p+A}$ for each $i$-th bin in $j$-th beam data such that the average of $f_{p+A}$ inside the bin matches the measured data $y_{ji}$: 
\begin{equation}
\label{eq:fitDiffModif}
f_{p+A} \to f_{p+A}' = f_{p+A} - \langle f_{p+A} \rangle_{ji} + y_{ji} \pm \sigma_{ji},
\end{equation}
where variables and notations are same as Eq.~(\ref{eq:9}). The sign of $\sigma_{ji}$ is positive if $\langle f_{p+A} \rangle_{ji} > y_{ji}$, otherwise it is negative. %We recalculated the $w$ in Eq.~\ref{eq:weight} with modification $f_{p+A} \to f_{p+A}'$ if the difference between data $y_{ji}$ and fit $\langle f_{p+A} \rangle_{ji}$ is larger than 1$\sigma_{ji}$. Writing just for notational consistency with other systematic uncertainty sources, this modification is equivalent to defining $\varepsilon$ in Eq.~(\ref{eq:11}) as:
The relative difference of the modified fit $f_{p+A}'$ from the default $f_{p+A}$ is defined as $\varepsilon$:
\begin{equation}
\label{eq:13}
 \varepsilon_5 =  \begin{cases}
    \frac{f_{p+A}'-f_{p+A}}{f_{p+A}} & \textrm{for bins where $|\langle f_{p+A} \rangle_{ji}-y_{ji}|>\sigma_{ji}$,} \\
    0                 & \textrm{otherwise}
  \end{cases}
\end{equation}

An example of $\varepsilon_{5}$ is shown in Fig.~\ref{fig:sys_fitFuncForm}.

\begin{figure}[!th]
\centering
\includegraphics[width=0.98\textwidth]{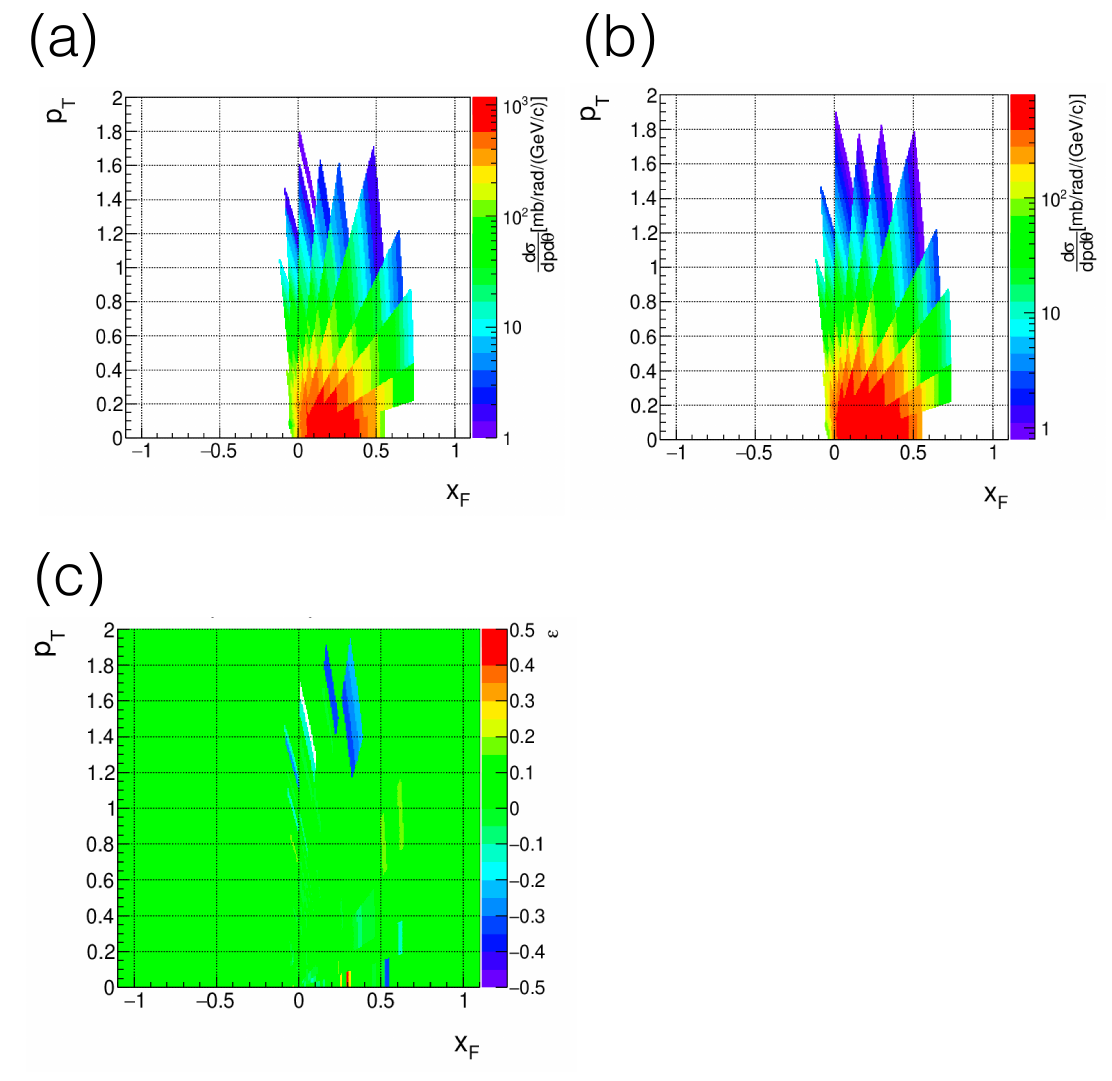}
 \caption{(a) Differential cross-sections of $\pi^+$ production measured in NA61, i.e., $y_{ij}$ in Eq.~\ref{eq:fitDiffModif}. (b) Our fit result to the NA61 data, i.e., $\langle f_{p+A} \rangle_{ji}$ in Eq.~\ref{eq:fitDiffModif}. (c) $\varepsilon$ for fitting $\chi^2$ (sys.5) at NA61 beam momentum.}
\label{fig:sys_fitFuncForm}
\end{figure}

\item Feynman scaling (Sys. 6 and 7)

We used data with beam momenta up to 450 GeV/$c$. For the higher momentum region, the weight for $p_{in} = 450$ GeV/$c$ was used assuming perfect Feynman scaling. However, the scaling is only valid in the limit where the momentum approaches infinity.
The uncertainty of the scaling was evaluated using the DPMJET-III event generator implemented in CRMC. The absolute scale of differential cross-section changes by $\sim$8\% between 400 and 1000 GeV/$c$, as shown in Fig.~\ref{fig:sys_FeynmanScale}(a). For $>$ 400 GeV/$c$, this difference is used as $\varepsilon_7$ in Eq.~(\ref{eq:11}). 
We also found the shape difference of differential cross-section distribution as shown in Fig.~\ref{fig:sys_FeynmanScale}(b). We considered this difference as $\varepsilon_6$ for the $>$ 158 GeV/$c$ region. 

\begin{figure}[!th]
\centering
\includegraphics[width=0.98\textwidth]{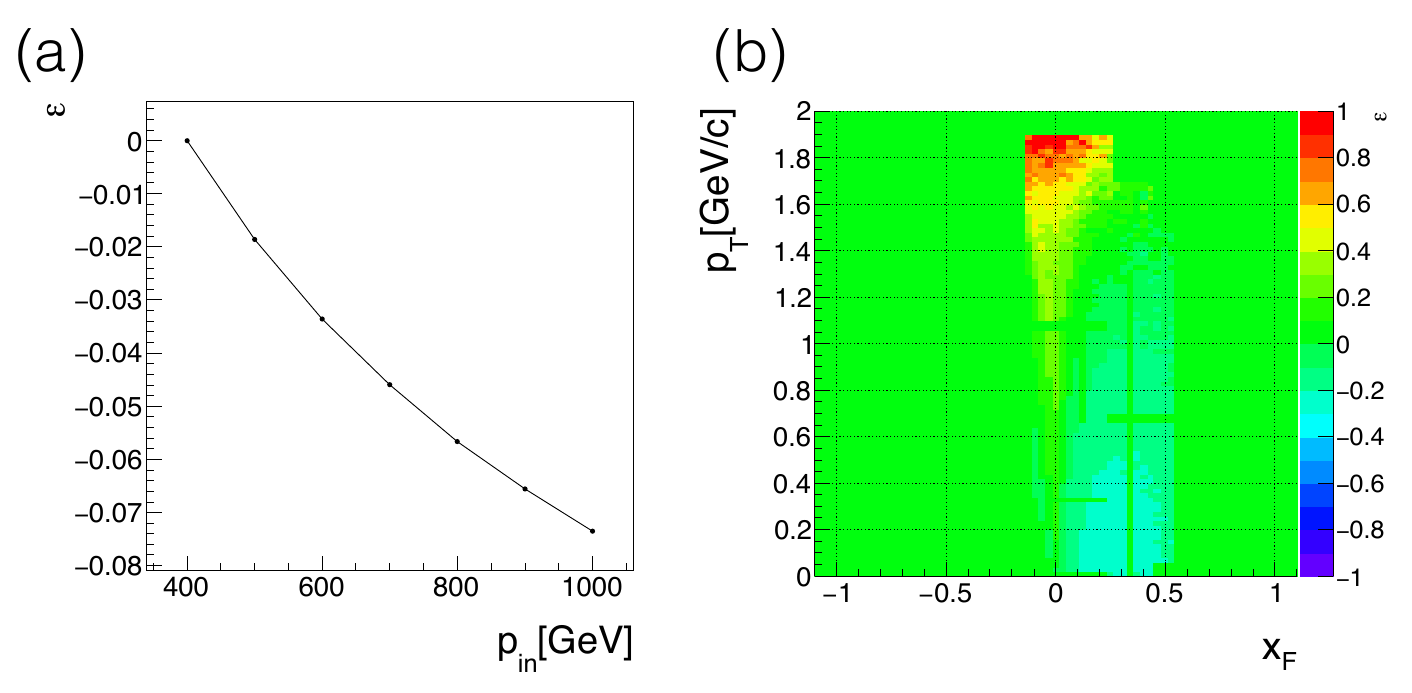}
 \caption{(a) Scaling factor of differential cross-section of $\pi^+$ production predicted by DPMJET-III. normalized to be 1 at $p_{in} = 400$ GeV/$c$. (b) Ratio of differential cross-section of $\pi^+$ production at $p_{in} = 1000$ GeV/$c$ to that at $p_{in} = 158$ GeV/$c$.}
\label{fig:sys_FeynmanScale}
\end{figure}

\item Uncertainty of A-dependence (Sys. 8, 9, and 10)

 In our fitting function, dependence on the target atom was described by $\alpha$ function (Eq.~(\ref{eq:7})). The parameters of the $\alpha$ were derived by fitting HARP Be, C, and Al data of $\pi^\pm$ and $p$ production, as described in Section~\ref{sec:parameterization}. The fitting error of $\alpha$ is considered a systematic uncertainty. Using error propagation, we calculated the error of $\left(A_{air} / A_{C} \right)^\alpha$ as shown in Fig.~\ref{fig:sys_AScaling}(a). The error is used as $\varepsilon_8$ in Eq.~(\ref{eq:11}). 

 We used $\alpha$ derived from HARP 12 GeV/$c$ data for a higher beam momentum region, assuming the momentum dependence of $\alpha$ is small. We also used $\alpha$ derived from $\pi^{+(-)}$ data for $K^{+(-)}$, assuming particle dependence of $\alpha$ is small. Such assumptions should be a systematic uncertainty source. We simulated $p+Air$ and $p+C$ interactions at $p_{in}=12$ GeV/$c$ and higher momenta using CRMC DPMJET-III. We then calculated the ratio of $p+Air$ differential cross-section to the $p+C$ differential cross-section. The difference of the ratio at $p_{in} > 12$ GeV/$c$ from the one at $p_{in} = 12$ GeV/$c$ was considered $\varepsilon_9$ for the uncertainty for momentum dependence. Similarly, we calculated the difference of the ratio between $\pi^{+,-}$ and $K^{+,-}$ productions, for the uncertainty for the particle dependence. An example is shown in Fig.~\ref{fig:sys_AScaling}(b). 

 The phase-space coverage of HARP data is smaller than those of NA61 and NA49 data. Thus, we needed to extrapolate the $\alpha$ function to the uncovered region, which should be a source of systematic uncertainty. We evaluated the differential cross-section ratio of $p+air$ to $p+C$ at $p_{in} = 12$ GeV using CRMC DPMJET-III. The ratio was then compared with $\left(A_{air} / A_{C} \right)^\alpha$ evaluated using the extrapolated $\alpha$, as shown in Fig.~\ref{fig:sys_AScaling}(c). The relative difference was used as $\varepsilon_{10}$.

\end{itemize}

\begin{figure}[!th]
\centering
\includegraphics[width=0.98\textwidth]{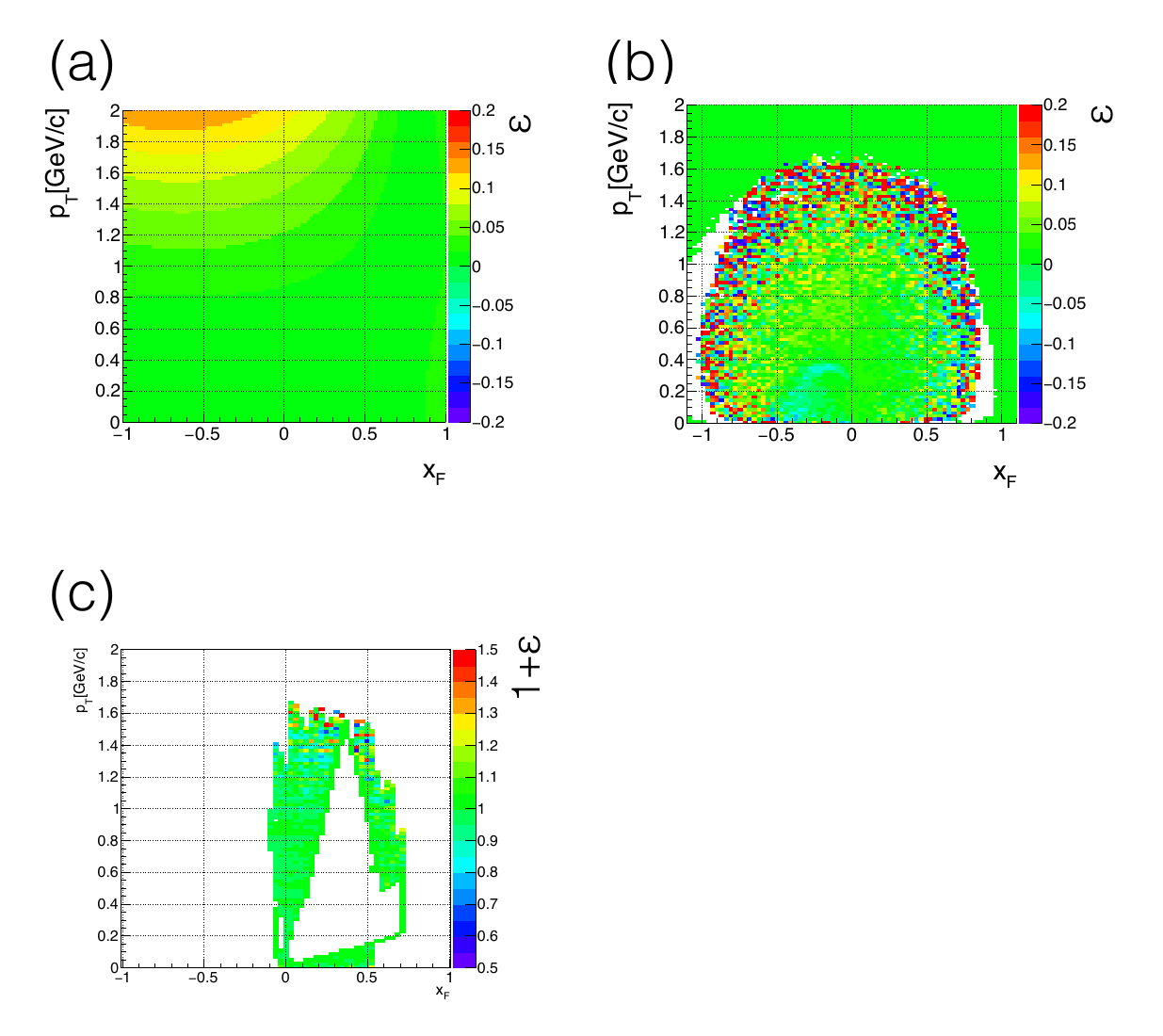}
 \caption{Systematic uncertainty related to $\alpha$ function. (a) Uncertainty related to fitting error for 12-GeV $\pi^+$ production. (b) Uncertainty related to the PID and $p_{in}$ dependence for $K^+$ production at $p_{in}=$ 31 GeV/$c$ . (c) Uncertainty related to uncovered phase space. }
\label{fig:sys_AScaling}
\end{figure}

\begin{figure}[!ht]
\centering
\subfigure[$\bar{\nu}_e$]{\includegraphics[width=0.45\textwidth]{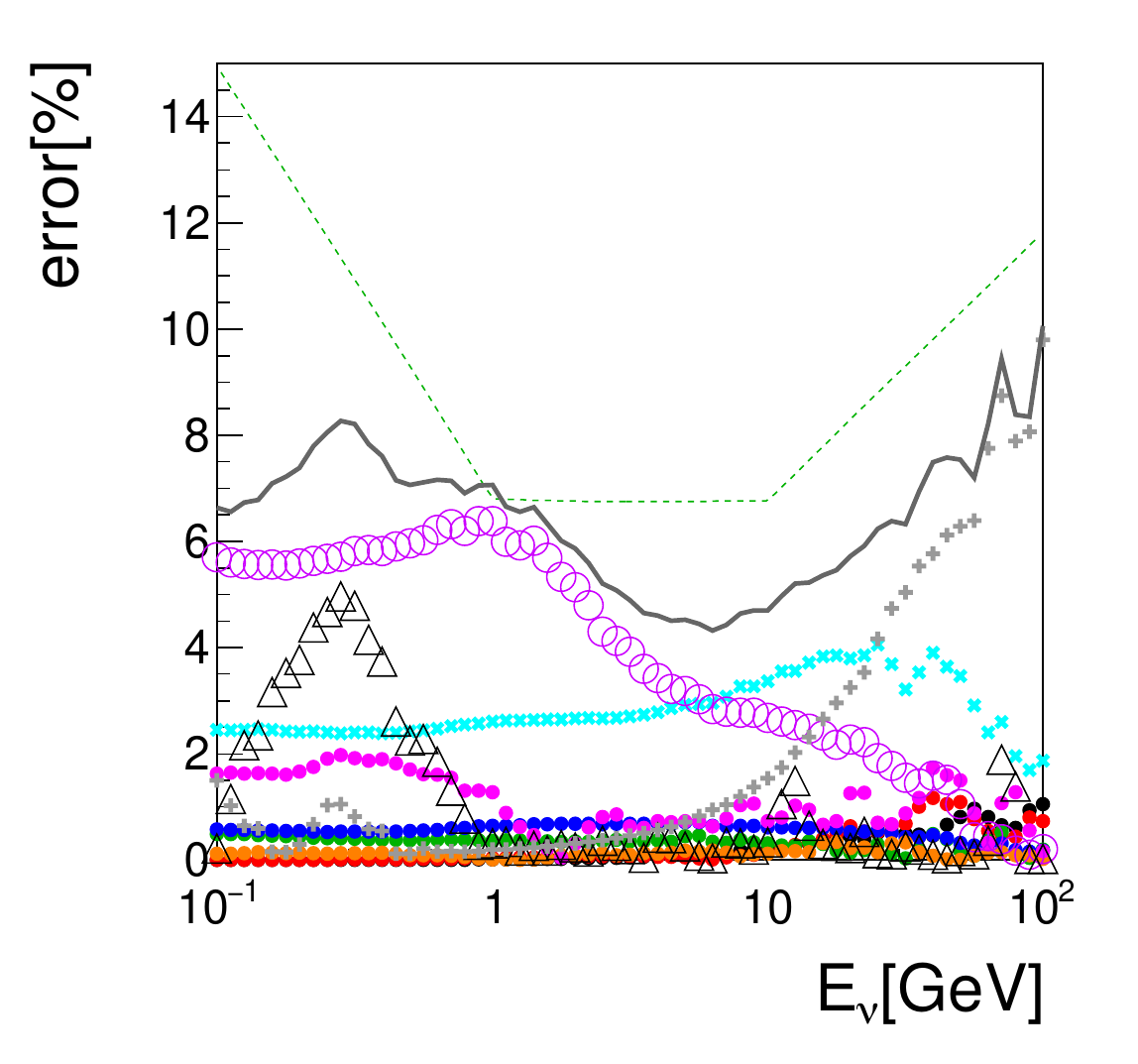}}
\subfigure[$\left(\nu_{\mu}-\bar{\nu}_{\mu}\right)/\left(\nu_e-\bar{\nu}_e\right)$]{\includegraphics[width=0.45\textwidth]{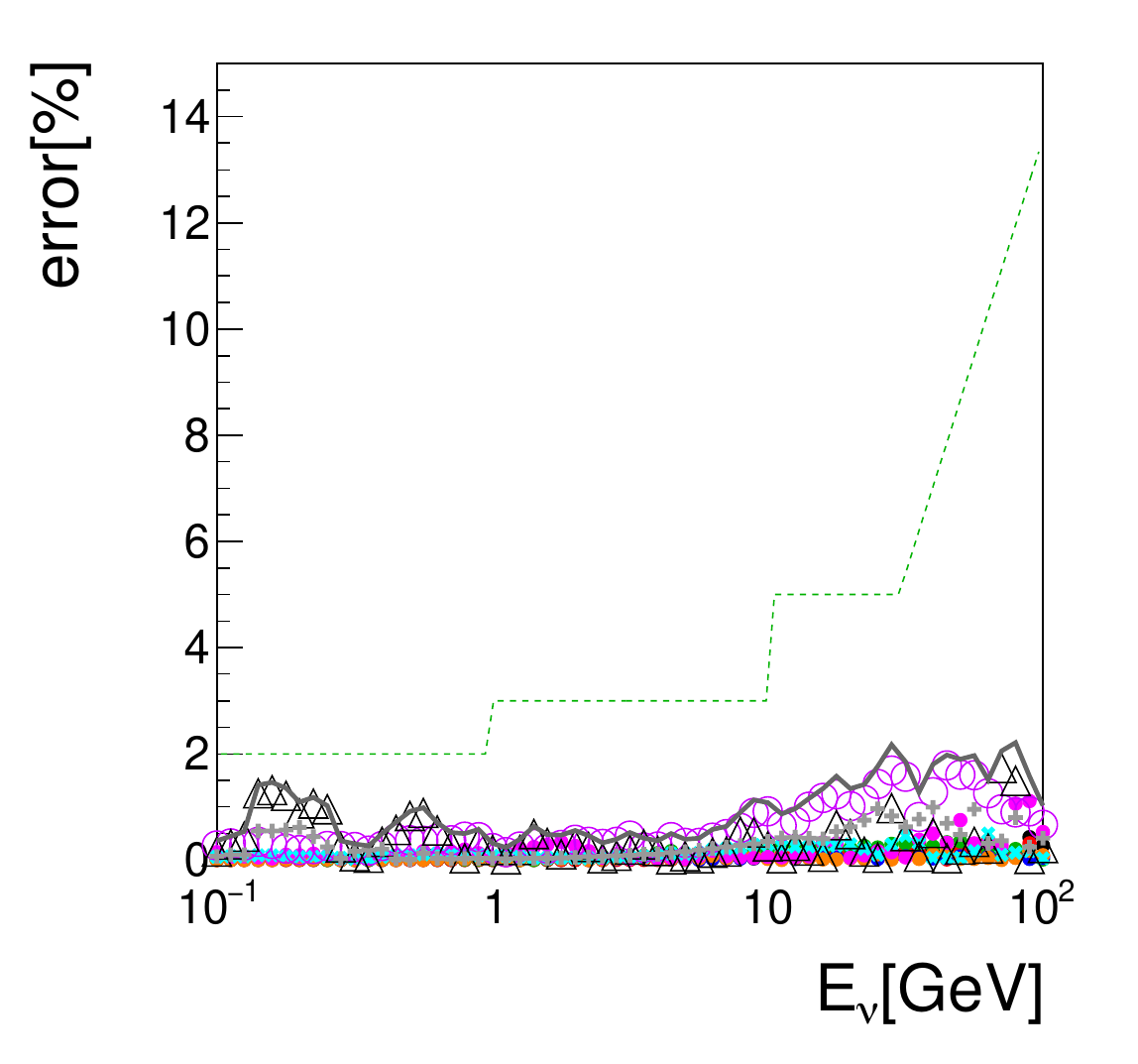}}
\\
\subfigure[$\bar{\nu}_\mu/\nu_{\mu}$]{\includegraphics[width=0.45\textwidth]{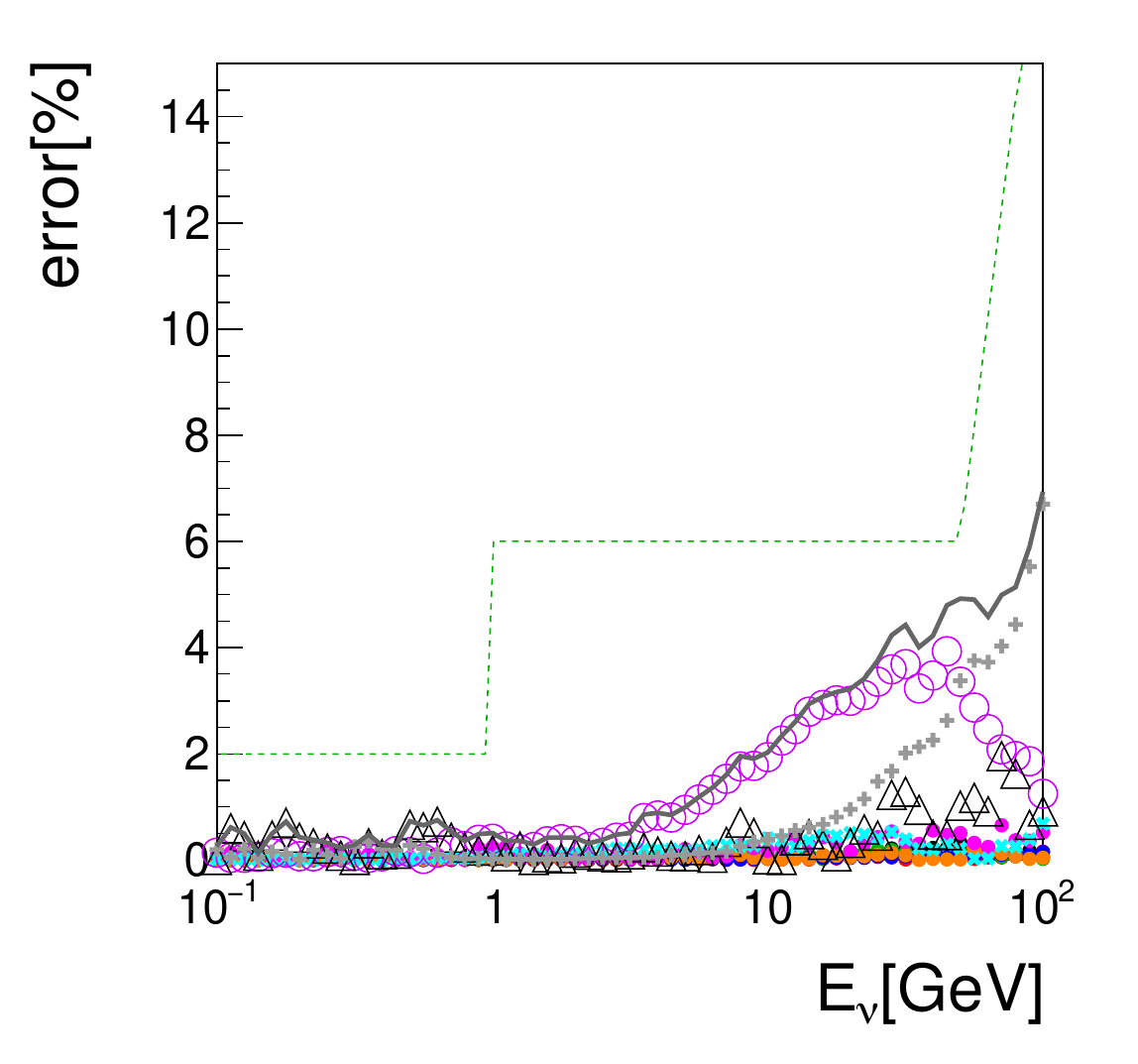}}
\subfigure[$\bar{\nu}_e/\nu_e$]{\includegraphics[width=0.45\textwidth]{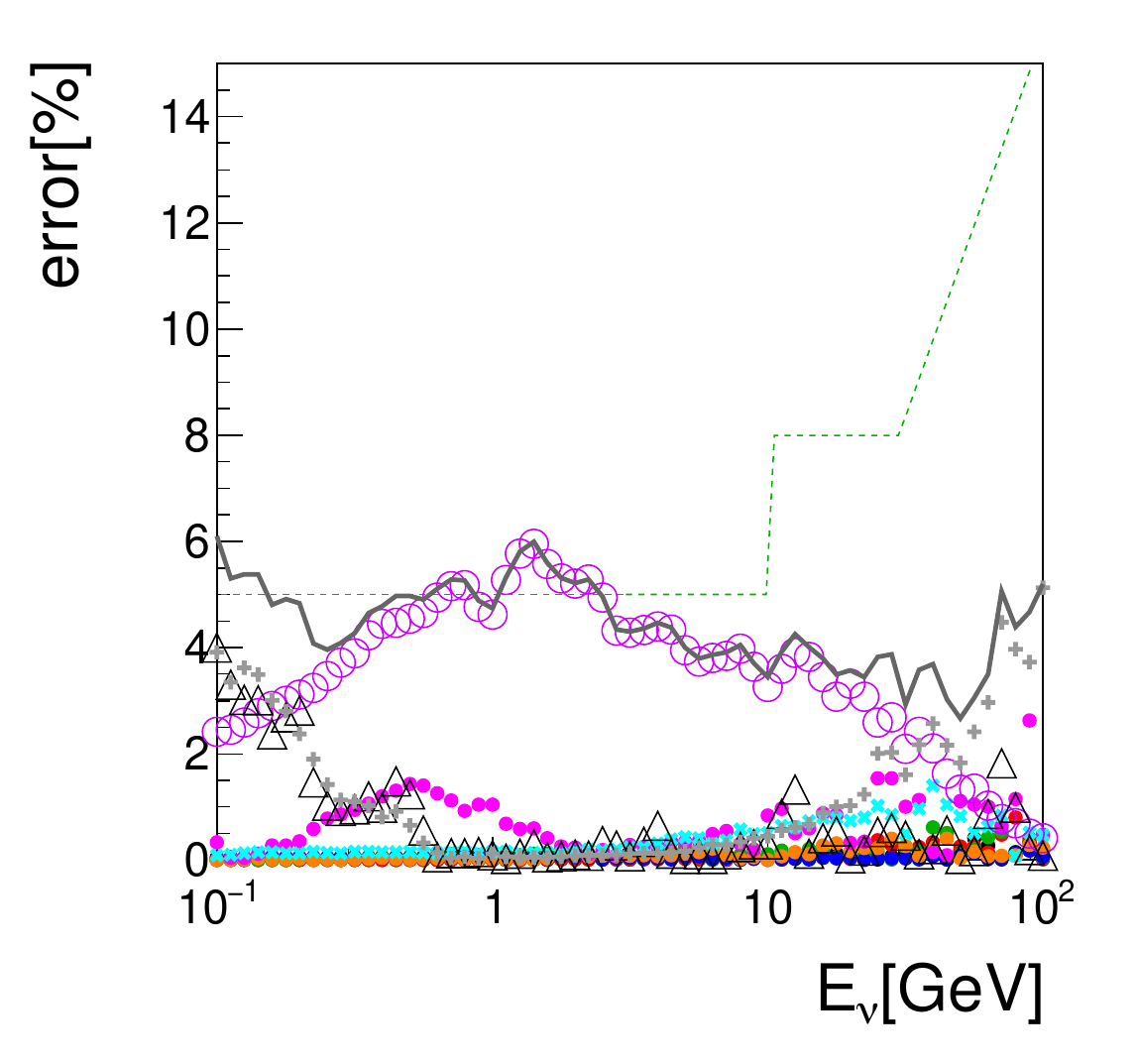}}
\caption{Systematic uncertainties in accelerator tuning for (a) $\bar{\nu}_e$ flux, (b) $(\nu_{\mu}+\bar{\nu}_{\mu})/(\nu_e+\bar{\nu}_e)$ ratio, (c) $\bar{\nu}_\mu/\nu_{\mu}$ ratio, and (d) $\bar{\nu}_e/\nu_{e}$. The gray line shows the total systematic uncertainty. The green dashed line shows the uncertainty evaluated for the muon tuning\cite{HKKM2006}. The markers show each uncertainty listed in Table~\ref{tab:Uncertainty}. Cyan $\times$ : measurement normalization (Sys.1), purple $\bigcirc$ : measurement error in bins (Sys.2), black $\triangle$ : phase space coverage for $3\le p_{in} \le 1000$ GeV/$c$ region(Sys.3), gray + : phase space coverage for $p_{in}<3$ GeV/$c$ and $p_{in}>1000$ GeV/$c$ regions(Sys.4), magenta dots : Sys.5, red dots : Sys.6, black dots : Sys.7, blue dots : Sys.8, green dots : Sys.9, orange dots : Sys.10. 
}
 \label{fig:sysSummary}
\end{figure}

The summaries of the systematic uncertainties evaluated above for the flux and flux ratio are shown in Fig.~\ref{fig:sysSummary}. The total systematic uncertainty was calculated by summing up all uncertainties in quadrature. It was evaluated to be $\sim$7\%--9\% in the $E_\nu < 1$ GeV region for the $\nu$ flux, whereas the muon tuning~\cite{MuTune,HKKM2006} could provide only a conservative uncertainty estimate in the $E_\nu < 1$ GeV region, owing to the lack of muon observations at low energies. Our new tuning contributed a reasonable and smaller uncertainty on that region.
In the energy region below a few GeV, the main source of uncertainty is the measurement errors in the accelerator data. In such low-energy regions, HARP and E910 provide the data, which have larger measurement errors compared with NA61 and NA49. Further reduction of the flux uncertainty is expected with future, more accurate accelerator measurements at $p_{beam} < 20$ GeV/$c$. %whereas the muon tuning~\cite{MuTune,HKKM2006} contributed only a conservative uncertainty of the flux in lower momentum region than $E_\nu < 1$ GeV/$c$ owing to the lack of $\mu$ observation. 
The uncertainty above 10 GeV is large because the beam energy of the data used in this study was limited to 450 GeV/$c$. Improvements can be expected by using higher-energy beam data or by extending the scope of the tuning to higher energies.

% \begin{figure}[t]
% \centering
% \includegraphics[width=0.98\textwidth]{img/sysSummary.pdf}
%  \caption{Systematic uncertianties in accelerator tuning for (a) $\nu_\mu$ flux, (b) $(\nu_{\mu}+\bar{\nu}_{\mu})/(\nu_e+\bar{\nu}_e)$ ratio, (c) $\bar{\nu}_\mu/\nu_{\mu}$ ratio, and (d) $\bar{\nu}_e/\nu_{e}$. The dots shows each uncertainty listed in Table~\ref{tab:Uncertainty}. The color meaning is also written in the table. The open circle shows the total systematic uncertainty. The solid line shows the uncertainty evaluated for the muon tunig\cite{HKKM2006}}
% \label{fig:sysSummary}
% \end{figure}

\section{Summary}
\label{sec:summary}

We implemented accelerator-data-driven tuning into our atmospheric neutrino flux MC simulation. Several hadron production measurements in beam experiments, whose beam momenta range from 3 to 450 GeV/$c$, have been used for the tuning. We used the weighting method to tune the difference of the differential cross-sections between data and MC. Although the neutrino flux prediction was reduced by 5\%--10\% with the tuning, the modified flux is in agreement with the previously published flux within its uncertainty. 
We evaluated the systematic uncertainties relevant to our accelerator tuning. We obtained a smaller and more reasonable uncertainty estimate for the neutrino momentum region below 1 GeV/$c$, where the conventional muon tuning could provide only a conservative error estimate. The main uncertainty arises from the measurement error of the accelerator data. Future, more accurate measurements at low beam momentum will improve the uncertainty. In this study, only beam data up to 450 GeV were used; therefore, the uncertainty becomes large in the $p_{\nu}>10$ GeV/$c$ region. Expanding the scope of the tuning to the high-energy side will reduce such uncertainty. %We placed a smaller and reasonable uncertainty in neutrino momentum lower than 1 GeV/$c$, where the conventional muon tuning presented a conservative error estimation only. 

The accelerator-data-driven tuning should complement the conventional muon tuning. The combined analysis of the accelerator tuning and the muon tuning is a topic for future study and will be able to further suppress the systematic uncertainty of the flux prediction. 

\section*{Acknowledgment}
We would like to thank L. Cook and G. Barr for the helpful discussion. 
This work was supported by the Japanese Ministry of Education, Culture, Sports, Science and Technology, Grant-in-Aid for Scientific Research, JSPS KAKENHI Grant No. 24K23938.

\let\doi\relax

\end{document}